\documentclass[10pt,journal]{IEEEtran}
\usepackage{xcolor,soul,framed,caption} 

\colorlet{shadecolor}{yellow}
\usepackage[pdftex]{graphicx}
\graphicspath{{../pdf/}{../jpeg/}}
\DeclareGraphicsExtensions{.pdf,.jpeg,.png}

\usepackage[cmex10]{amsmath}

\usepackage{array}
\usepackage{mdwmath}
\usepackage{mdwtab}
\usepackage{eqparbox}
\usepackage{url}
\usepackage{supertabular}                        

\usepackage{hyperref}
\hypersetup{
  colorlinks=true,
  linkcolor=blue,
  urlcolor=blue,
  citecolor=blue
}  
\usepackage{amsmath,amssymb}
\usepackage{caption}
\usepackage{subcaption}
\usepackage{float}
\usepackage{booktabs}
\usepackage{multirow}
\usepackage{graphicx}
\usepackage{color}
\usepackage{tabularray}
\usepackage{hhline}
\usepackage{wasysym}
\usepackage[ruled, lined, longend, linesnumbered]{algorithm2e}
\usepackage{algorithmic}
\usepackage{tikz}
\usetikzlibrary{positioning,fit,shapes.misc,arrows.meta,calc}

\usepackage{setspace}
\usepackage{titlesec}
\titlespacing*{\section}{0pt}{0.5\baselineskip}{0.3\baselineskip}
\titlespacing*{\subsection}{0pt}{0.4\baselineskip}{0.2\baselineskip}
\titlespacing*{\subsubsection}{0pt}{0.3\baselineskip}{0.1\baselineskip}

\renewcommand{\thesection}{\arabic{section}}
\renewcommand{\thesubsection}{\thesection.\arabic{subsection}}
\renewcommand{\thesubsubsection}{\thesubsection.\arabic{subsubsection}}

\titleformat{\section}
  {\bfseries}{\thesection}{1em}{}
\titleformat{\subsection}
  {\bfseries}{\thesubsection}{1em}{}
\titleformat{\subsubsection}
  {\bfseries}{\thesubsubsection}{1em}{}
  
\makeatletter
\def\@IEEENORMtitlevspace{1\baselineskip}
\def\@IEEEMINtitlevspace{0.3\baselineskip}  
\makeatother


\begin{document}
\bstctlcite{IEEEexample:BSTcontrol}
    \title{A Knowledge Distillation-empowered Adaptive Federated Reinforcement Learning Framework for Multi-Domain IoT Applications Scheduling}
  \author{Zhiyu Wang, Mohammad Goudarzi, Mingming Gong, and Rajkumar Buyya
  \thanks{Zhiyu Wang and Rajkumar Buyya are with The Quantum Cloud Computing and Distributed Systems (qCLOUDS) Laboratory, School of Computing and Information Systems, The University of Melbourne, Australia (e-mail: zhiywang1@student.unimelb.edu.au, rbuyya@unimelb.edu.au).}
  \thanks{Mingming Gong is with School of Mathematics and Statistics, The University of Melbourne, Australia (email: mingming.gong@unimelb.edu.au)}
  \thanks{Mohammad Goudarzi is with The Faculty of Information Technology, Monash University, Australia (email: mohammad.goudarzi@monash.edu)}
}  

\maketitle
\IEEEaftertitletext{\vspace{-1.5em}}

\begin{abstract}
The rapid proliferation of Internet of Things (IoT) applications across heterogeneous Cloud-Edge-IoT environments presents significant challenges in distributed scheduling optimization. Existing approaches face issues, including fixed neural network architectures that are incompatible with computational heterogeneity, non-Independent and Identically Distributed (non-IID) data distributions across IoT scheduling domains, and insufficient cross-domain collaboration mechanisms. This paper proposes KD-AFRL, a Knowledge Distillation-empowered Adaptive Federated Reinforcement Learning framework that addresses multi-domain IoT application scheduling through three core innovations. First, we develop a resource-aware hybrid architecture generation mechanism that creates dual-zone neural networks enabling heterogeneous devices to participate in collaborative learning while maintaining optimal resource utilization. Second, we propose a privacy-preserving environment-clustered federated learning approach that utilizes differential privacy and K-means clustering to address non-IID challenges and facilitate effective collaboration among compatible domains. Third, we introduce an environment-oriented cross-architecture knowledge distillation mechanism that enables efficient knowledge transfer between heterogeneous models through temperature-regulated soft targets. Comprehensive experiments with real Cloud-Edge-IoT infrastructure demonstrate KD-AFRL's effectiveness using diverse IoT applications. Results show significant improvements over the best baseline, with 21\% faster convergence and 15.7\%, 10.8\%, and 13.9\% performance gains in completion time, energy consumption, and weighted cost, respectively. Scalability experiments reveal that KD-AFRL achieves 3-5 times better performance retention compared to existing solutions as the number of domains increases.

\end{abstract}


\IEEEpeerreviewmaketitle

\section{Introduction}
The rapid proliferation of Internet of Things (IoT) applications has fundamentally transformed computing paradigms, creating unprecedented demands for intelligent resource management in heterogeneous Cloud-Edge-IoT environments \cite{yan2024edge}. Modern IoT deployments span multiple autonomous domains—from smart cities and industrial automation to healthcare monitoring and autonomous vehicles—each exhibiting distinct computational capabilities, workload characteristics, and operational constraints \cite{andreoli2025multi, zhang2025gai}. These applications typically consist of interdependent tasks forming complex Directed Acyclic Graphs (DAGs), requiring sophisticated scheduling strategies to optimize completion time, energy consumption, and operational costs while respecting dependency constraints and resource limitations \cite{cong2025task}. For example, in a multi-city smart transportation collaboration scenario, traffic management departments across different cities need to schedule various IoT applications such as video surveillance analysis, traffic flow prediction, and signal optimization, but each city faces distinctly different environmental characteristics: some cities have relatively stable traffic patterns and good network conditions, others face highly dynamic traffic flows and unstable network environments, while still others need to handle scheduling challenges under extreme weather conditions. By leveraging multi-domain collaborative learning, cities with diverse operational contexts can exchange and incorporate specialized scheduling insights, leading to more resilient, adaptive, and globally optimized IoT application performance across heterogeneous environments.

To address these complex scheduling optimization challenges, Deep Reinforcement Learning (DRL) has emerged as a promising solution for adaptive policy learning \cite{zhu2023transfer}. However, traditional centralized DRL scheduling faces significant scalability and adaptability challenges, struggling to capture the dynamic nature of multi-domain IoT ecosystems where resource availability, network conditions, and workload patterns fluctuate continuously across distributed domains \cite{wang2025tf}. These limitations become particularly pronounced when managing heterogeneous computational resources ranging from resource-constrained IoT devices to high-performance cloud servers. To overcome these limitations, distributed DRL has emerged to improve system scalability by distributing computation across multiple devices. However, existing distributed DRL scheduling methods suffer from several fundamental deficiencies. First, most methods employ fixed neural network architectures that cannot adapt to computational heterogeneity, resulting in resource mismatches across devices. Second, existing works operate under unrealistic IID data assumptions, neglecting the non-IID nature of real-world multi-domain deployments where different domains exhibit distinct environmental characteristics and workload patterns. Third, current distributed DRL methods lack effective cross-domain collaboration mechanisms, failing to exploit the potential for knowledge sharing between different domains.

Federated Learning (FL) offers a compelling solution for enabling collaborative learning across distributed domains while preserving data locality and privacy \cite{chellapandi2023federated}. By allowing multiple domains to jointly train machine learning models without sharing raw data, FL addresses privacy concerns and reduces communication overhead associated with centralized approaches \cite{liu2024vertical}. However, applying federated learning to multi-domain IoT scheduling introduces several fundamental challenges. First, the heterogeneity in computational capabilities across domains necessitates adaptive model architectures that can scale appropriately to device constraints while maintaining learning effectiveness \cite{wang2025empowering}. Second, the non-IID nature of scheduling environments across domains can significantly degrade federated learning performance when domains with dissimilar characteristics attempt to share model parameters directly \cite{lu2024federated}. Third, domains with different computational capabilities often employ architectures of varying complexity, making direct parameter aggregation impossible and preventing resource-constrained domains from benefiting from knowledge acquired by more capable domains \cite{fan2025ten}.

To address these challenges, we propose KD-AFRL, a Knowledge Distillation-empowered Adaptive Federated Reinforcement Learning framework that enables effective multi-domain IoT application scheduling. By introducing resource-aware adaptive architecture generation, privacy-preserving environment-clustered federated learning, and environment-oriented cross-architecture knowledge distillation, KD-AFRL enables heterogeneous devices, from resource-constrained IoT devices to powerful cloud servers, to collaboratively learn optimal scheduling policies despite non-IID data distributions across domains, while preserving data privacy and adapting to their computational constraints.

The main contributions of this work are fourfold:
\begin{itemize}
\item We design a resource-aware hybrid architecture generation mechanism that dynamically adapts model complexity to each device's computational capacity. Leveraging a dual-zone architecture that comprises shared foundational zones and personalized adaptation zones, the mechanism preserves federated learning compatibility across domains, enabling heterogeneous devices to participate in collaborative learning while maximizing optimal resource utilization.

\item We propose a privacy-preserving environment-clustered federated learning mechanism that addresses non-IID challenges in multi-domain deployments. The mechanism leverages K-means clustering to group domains with similar environmental characteristics, facilitating targeted collaboration among compatible domains while mitigating negative transfer from dissimilar environments. To ensure data confidentiality, we incorporate $\epsilon$-differential privacy throughout the clustering and federated aggregation pipeline, protecting sensitive operational information without compromising learning effectiveness.

\item We propose an environment-oriented cross-architecture knowledge distillation mechanism that enables efficient knowledge transfer between heterogeneous models based on environmental similarities. Through temperature-regulated soft targets, the mechanism allows small models on resource-constrained devices to achieve competitive performance compared to large models on high-end devices. 

\item We conduct comprehensive practical evaluation across distributed scheduling domains with real Cloud-Edge-IoT infrastructure, demonstrating the effectiveness and scalability of KD-AFRL using diverse real-world IoT applications spanning different computational characteristics and resource requirements.
\end{itemize}

The rest of the paper is organized as follows. Section~\ref{related_work} reviews related work. Section~\ref{system_model} presents the system model and problem formulation. Section~\ref{kd-afrl} details the KD-AFRL framework. Section~\ref{evaluation} presents experimental evaluation. Section~\ref{conclusions} concludes the paper.

\section{Related Work}
\label{related_work}
In this section, we review existing DRL techniques for IoT scheduling, categorizing them into centralized and distributed approaches, and identify research gaps through qualitative comparison.

\subsection{Centralized DRL for IoT Scheduling}
Tang et al. \cite{tang2025adaptive} proposed RASO based on Deep Q-Network (DQN) for collaborative task offloading in Mobile Edge Computing (MEC) networks. The method employs spatial indexing and fine-grained task recombination to minimize offloading delay and energy consumption. Zhu et al. \cite{zhu2025drl} proposed a Proximal Policy Optimization (PPO)-based approach with hybrid actor-critic networks for joint wireless charging and computation offloading in wireless-powered multi-access edge computing (WP-MEC). The objective is to maximize utility characterized by wireless devices' residual energy and social relationship strength. Fan et al. \cite{fan2025vehicular} proposed a Softmax Deep Double Deterministic Policy Gradients (DDPG)-based resource orchestration scheme for vehicle collaborative networks. The method aims at minimizing total cost involving latency and energy consumption. Chen et al. \cite{chen2024real} proposed DODQ based on DQN for cloud-edge computing environments. The approach models mobile applications as DAGs to adaptively handle dynamic resource changes and parallel task scheduling without presetting task priorities. Wang et al. \cite{wang2024deep} proposed DRLIS based on PPO for IoT application scheduling in heterogeneous edge/fog computing environments. The method optimizes response time and load balancing for DAG-based applications. Hsieh et al. \cite{hsieh2023deep} investigated the task assignment problem in cooperative MEC networks, developing and comparing Double-DQN, Policy Gradient, and Actor-Critic algorithms for task optimization. The results demonstrated that the Actor-Critic approach performed best in optimizing delay under dynamic MEC environments. Zhao et al. \cite{zhao2023meson} proposed MESON based on DDPG for urban vehicular edge computing. The scheme incorporates vehicle mobility detection and task priority determination to minimize average response time and energy consumption. Chi et al. \cite{chi2024task} proposed a scheme that combines Double Dueling DQN (D3QN) and prioritized experience for task offloading in edge-assisted Industrial Internet of Things (IIoT). The scheme reduces average task cost and improves task completion rate by enhancing action selection accuracy and convergence speed. 

\subsection{Distributed DRL for IoT Scheduling}
Wu et al. \cite{wu2024proactive} proposed a distributed DQN-based algorithm with temporal convolution sequence network (TCSN) for proactive caching in 6G cloud-edge collaboration computing. The distributed approach maximizes edge hit ratio while minimizing content access latency and traffic cost. Zhao et al. \cite{zhao2024asynchronous} proposed ADTO, an Asynchronous Advantage Actor-Critic (A3C)-based solution for multi-hop task offloading in RSU-assisted Internet of Vehicles (IoV) networks. The approach establishes mobility models and forwarding vehicle selection mechanisms to minimize task delay. Wang et al. \cite{wang2025tf} proposed a Transformer-enhanced Distributed DRL technique (TF-DDRL) based on Importance Weighted Actor-Learner Architectures (IMPALA) for scheduling heterogeneous IoT applications in edge and cloud environments. The approach incorporates prioritized experience replay and off-policy correction to reduce response time, energy consumption, and monetary cost. Zhou et al. \cite{zhou2023cost} proposed DRLCOSCM using A3C algorithm for three-tier mobile cloud-edge computing. The approach minimizes cloud service cost while meeting delay requirements of mobile users. Zhang et al. \cite{zhang2024lsia3cs} proposed LsiA3CS based on A3C for task scheduling in IIoT. The approach incorporates Markov game modeling and heuristic guidance to reduce task completion times. Ju et al. \cite{ju2023noma} proposed an A3C-based energy-efficiency secure offloading (EESO) scheme for vehicular edge computing networks. The approach aims at minimizing system energy consumption while ensuring security. Liu et al. \cite{liu2023asynchronous} proposed an A3C-based algorithm for collaborative task computing and on-demand resource allocation in vehicular edge computing. The approach maximizes system utility through optimal task and resource scheduling policy considering service migration and available vehicle resources. Shen et al. \cite{shen2024asynchronous} proposed AFO, an asynchronous federated PPO-based task offloading algorithm for dependency-aware UAV-assisted vehicular networks. The approach enhances data diversity to minimize average task execution delay and energy consumption. 

\subsection{A Qualitative Comparison}
To systematically analyze the existing literature and identify research gaps, we conduct a comprehensive qualitative comparison of related works presented in Table~\ref{tab:related_works}, evaluating them across four critical dimensions: application properties, system properties, technique properties, and evaluation methodology.

\subsubsection{Comparative Analysis Dimensions}
Application Properties evaluate workload complexity through task number and dependency. System Properties assess infrastructure scope including application types, computing environments, heterogeneity, and multi-domain support. Technique Properties analyzes algorithmic approaches (centralized vs. distributed), DRL techniques, optimization objectives, and adaptive architecture support. Evaluation Methodology distinguishes simulation-based from practical deployment evaluation.

\subsubsection{Research Gap Identification}
Based on our systematic analysis, we identify four fundamental research gaps that existing literature fails to address:

\textbf{Gap 1 - Limited Application Realism}: Only 5 works support task dependencies, and merely 2 evaluate real IoT applications, indicating a substantial disconnect between research assumptions and practical deployment requirements. Real-world IoT applications typically exhibit complex interdependencies and diverse computational characteristics that are not captured in simplified single-task or synthetic workload scenarios employed by most existing works.

\textbf{Gap 2 - Multi-Domain Coordination}: All existing works lack multi-domain support, despite real-world IoT deployments spanning multiple autonomous domains. This design limitation results in isolated scheduling systems, missing opportunities for cross-domain resource optimization and collaborative decision-making.

\textbf{Gap 3 - Adaptive Architecture Generation}: All existing works employ fixed neural network architectures, completely ignoring the substantial computational heterogeneity from IoT devices to cloud servers. This architectural rigidity leads to either severe resource underutilization on high-performance devices or computational overload on resource-constrained devices. 

\textbf{Gap 4 - Evaluation Limitations}: 14 works rely solely on simulation-based evaluation, lacking practical deployment verification and failing to capture real-world operational complexities and uncertainties. 

These identified gaps collectively motivate the development of KD-AFRL, which systematically addresses the core challenges of heterogeneous multi-domain IoT scheduling through resource-aware adaptive architecture generation, privacy-preserving environment-clustered federated learning, and environment-oriented cross-architecture knowledge distillation.

\renewcommand{\arraystretch}{1.5}
\begin{table*}[]
\centering
\caption{A qualitative comparison of our work with existing related works}
\label{tab:related_works}
\resizebox{\textwidth}{!}{%
\begin{tabular}{ccccccccccccccc}
\hline
\multicolumn{1}{|c|}{\multirow{3}{*}{Work}}  & \multicolumn{2}{c|}{Application Properties} & \multicolumn{5}{c|}{System Properties} & \multicolumn{6}{c|}{Technique Properties} & \multicolumn{1}{c|}{\multirow{3}{*}{Evaluation}} \\ \cline{2-14}
\multicolumn{1}{|c|}{} & \multicolumn{1}{c|}{\multirow{2}{*}{Task Number}} & \multicolumn{1}{c|}{\multirow{2}{*}{Dependency}} & \multicolumn{2}{c|}{IoT Device Layer} & \multicolumn{2}{c|}{Edge/Cloud Layer} & \multicolumn{1}{c|}{\multirow{2}{*}{Multi-Domain}} & \multicolumn{2}{c|}{Main Technique} & \multicolumn{3}{c|}{Optimization Objectives} & \multicolumn{1}{c|}{\multirow{2}{*}{\begin{tabular}[c]{@{}c@{}}Resource-aware\\Adaptive Architecture\end{tabular}}} & \multicolumn{1}{c|}{} \\ \cline{4-7} \cline{9-10} \cline{11-13}
\multicolumn{1}{|c|}{} & \multicolumn{1}{c|}{} & \multicolumn{1}{c|}{} & \multicolumn{1}{c|}{Real Applications} & \multicolumn{1}{c|}{Request Type} & \multicolumn{1}{c|}{Computing Environment} & \multicolumn{1}{c|}{Heterogeneity} & \multicolumn{1}{c|}{} & \multicolumn{1}{c|}{Type} & \multicolumn{1}{c|}{Algorithm} & \multicolumn{1}{c|}{Time} & \multicolumn{1}{c|}{Energy} & \multicolumn{1}{c|}{Multi Objective} & \multicolumn{1}{c|}{} & \multicolumn{1}{c|}{} \\ \hline

\multicolumn{1}{|c|}{Tang et al. \cite{tang2025adaptive}} & \multicolumn{1}{c|}{Multiple} & \multicolumn{1}{c|}{Independent} & \multicolumn{1}{c|}{\LEFTcircle} & \multicolumn{1}{c|}{Homogeneous} & \multicolumn{1}{c|}{Edge} & \multicolumn{1}{c|}{Heterogeneous} & \multicolumn{1}{c|}{$\times$} & \multicolumn{1}{c|}{\multirow{9}{*}{Centralized}} & \multicolumn{1}{c|}{DQN} & \multicolumn{1}{c|}{\checkmark} & \multicolumn{1}{c|}{\checkmark} & \multicolumn{1}{c|}{\checkmark} & \multicolumn{1}{c|}{$\times$} & \multicolumn{1}{c|}{Simulation} \\ \cline{1-8} \cline{10-15}

\multicolumn{1}{|c|}{Zhu et al. \cite{zhu2025drl}} & \multicolumn{1}{c|}{Single} & \multicolumn{1}{c|}{Independent} & \multicolumn{1}{c|}{\Circle} & \multicolumn{1}{c|}{Homogeneous} & \multicolumn{1}{c|}{Edge} & \multicolumn{1}{c|}{Homogeneous} & \multicolumn{1}{c|}{$\times$} & \multicolumn{1}{c|}{} & \multicolumn{1}{c|}{PPO} & \multicolumn{1}{c|}{$\times$} & \multicolumn{1}{c|}{\checkmark} & \multicolumn{1}{c|}{\checkmark} & \multicolumn{1}{c|}{$\times$} & \multicolumn{1}{c|}{Simulation} \\ \cline{1-8} \cline{10-15}

\multicolumn{1}{|c|}{Fan et al. \cite{fan2025vehicular}} & \multicolumn{1}{c|}{Single} & \multicolumn{1}{c|}{Independent} & \multicolumn{1}{c|}{\LEFTcircle} & \multicolumn{1}{c|}{Homogeneous} & \multicolumn{1}{c|}{Edge} & \multicolumn{1}{c|}{Heterogeneous} & \multicolumn{1}{c|}{$\times$} & \multicolumn{1}{c|}{} & \multicolumn{1}{c|}{DDPG} & \multicolumn{1}{c|}{\checkmark} & \multicolumn{1}{c|}{\checkmark} & \multicolumn{1}{c|}{\checkmark} & \multicolumn{1}{c|}{$\times$} & \multicolumn{1}{c|}{Simulation} \\ \cline{1-8} \cline{10-15}

\multicolumn{1}{|c|}{Chen et al. \cite{chen2024real}} & \multicolumn{1}{c|}{Multiple} & \multicolumn{1}{c|}{Dependent} & \multicolumn{1}{c|}{\LEFTcircle} & \multicolumn{1}{c|}{Heterogeneous} & \multicolumn{1}{c|}{Edge and Cloud} & \multicolumn{1}{c|}{Heterogeneous} & \multicolumn{1}{c|}{$\times$} & \multicolumn{1}{c|}{} & \multicolumn{1}{c|}{DQN} & \multicolumn{1}{c|}{\checkmark} & \multicolumn{1}{c|}{$\times$} & \multicolumn{1}{c|}{$\times$} & \multicolumn{1}{c|}{$\times$} & \multicolumn{1}{c|}{Simulation} \\ \cline{1-8} \cline{10-15}

\multicolumn{1}{|c|}{Wang et al. \cite{wang2024deep}} & \multicolumn{1}{c|}{Multiple} & \multicolumn{1}{c|}{Dependent} & \multicolumn{1}{c|}{\CIRCLE} & \multicolumn{1}{c|}{Heterogeneous} & \multicolumn{1}{c|}{Edge and Cloud} & \multicolumn{1}{c|}{Heterogeneous} & \multicolumn{1}{c|}{$\times$} & \multicolumn{1}{c|}{} & \multicolumn{1}{c|}{PPO} & \multicolumn{1}{c|}{\checkmark} & \multicolumn{1}{c|}{$\times$} & \multicolumn{1}{c|}{\checkmark} & \multicolumn{1}{c|}{$\times$} & \multicolumn{1}{c|}{Practical} \\ \cline{1-8} \cline{10-15}

\multicolumn{1}{|c|}{Hsieh et al. \cite{hsieh2023deep}} & \multicolumn{1}{c|}{Single} & \multicolumn{1}{c|}{Independent} & \multicolumn{1}{c|}{\LEFTcircle} & \multicolumn{1}{c|}{Heterogeneous} & \multicolumn{1}{c|}{Edge and Cloud} & \multicolumn{1}{c|}{Heterogeneous} & \multicolumn{1}{c|}{$\times$} & \multicolumn{1}{c|}{} & \multicolumn{1}{c|}{Actor-Critic} & \multicolumn{1}{c|}{\checkmark} & \multicolumn{1}{c|}{$\times$} & \multicolumn{1}{c|}{$\times$} & \multicolumn{1}{c|}{$\times$} & \multicolumn{1}{c|}{Simulation} \\ \cline{1-8} \cline{10-15}

\multicolumn{1}{|c|}{Zhao et al. \cite{zhao2023meson}} & \multicolumn{1}{c|}{Multiple} & \multicolumn{1}{c|}{Dependent} & \multicolumn{1}{c|}{\LEFTcircle} & \multicolumn{1}{c|}{Heterogeneous} & \multicolumn{1}{c|}{Edge} & \multicolumn{1}{c|}{Heterogeneous} & \multicolumn{1}{c|}{$\times$} & \multicolumn{1}{c|}{} & \multicolumn{1}{c|}{DDPG} & \multicolumn{1}{c|}{\checkmark} & \multicolumn{1}{c|}{\checkmark} & \multicolumn{1}{c|}{\checkmark} & \multicolumn{1}{c|}{$\times$} & \multicolumn{1}{c|}{Simulation} \\ \cline{1-8} \cline{10-15}

\multicolumn{1}{|c|}{Chi et al. \cite{chi2024task}} & \multicolumn{1}{c|}{Single} & \multicolumn{1}{c|}{Independent} & \multicolumn{1}{c|}{\Circle} & \multicolumn{1}{c|}{Homogeneous} & \multicolumn{1}{c|}{Edge} & \multicolumn{1}{c|}{Homogeneous} & \multicolumn{1}{c|}{$\times$} & \multicolumn{1}{c|}{} & \multicolumn{1}{c|}{DQN} & \multicolumn{1}{c|}{\checkmark} & \multicolumn{1}{c|}{\checkmark} & \multicolumn{1}{c|}{\checkmark} & \multicolumn{1}{c|}{$\times$} & \multicolumn{1}{c|}{Simulation} \\ \hline

\multicolumn{1}{|c|}{Wu et al. \cite{wu2024proactive}} & \multicolumn{1}{c|}{Single} & \multicolumn{1}{c|}{Independent} & \multicolumn{1}{c|}{\LEFTcircle} & \multicolumn{1}{c|}{Homogeneous} & \multicolumn{1}{c|}{Edge and Cloud} & \multicolumn{1}{c|}{Homogeneous} & \multicolumn{1}{c|}{$\times$} & \multicolumn{1}{c|}{\multirow{9}{*}{Distributed}} & \multicolumn{1}{c|}{DQN} & \multicolumn{1}{c|}{\checkmark} & \multicolumn{1}{c|}{$\times$} & \multicolumn{1}{c|}{\checkmark} & \multicolumn{1}{c|}{$\times$} & \multicolumn{1}{c|}{Simulation} \\ \cline{1-8} \cline{10-15}

\multicolumn{1}{|c|}{Zhao et al. \cite{zhao2024asynchronous}} & \multicolumn{1}{c|}{Single} & \multicolumn{1}{c|}{Independent} & \multicolumn{1}{c|}{\Circle} & \multicolumn{1}{c|}{Homogeneous} & \multicolumn{1}{c|}{Edge and Cloud} & \multicolumn{1}{c|}{Homogeneous} & \multicolumn{1}{c|}{$\times$} & \multicolumn{1}{c|}{} & \multicolumn{1}{c|}{A3C} & \multicolumn{1}{c|}{\checkmark} & \multicolumn{1}{c|}{$\times$} & \multicolumn{1}{c|}{$\times$} & \multicolumn{1}{c|}{$\times$} & \multicolumn{1}{c|}{Simulation} \\ \cline{1-8} \cline{10-15}

\multicolumn{1}{|c|}{Wang et al. \cite{wang2025tf}} & \multicolumn{1}{c|}{Multiple} & \multicolumn{1}{c|}{Dependent} & \multicolumn{1}{c|}{\CIRCLE} & \multicolumn{1}{c|}{Heterogeneous} & \multicolumn{1}{c|}{Edge and Cloud} & \multicolumn{1}{c|}{Heterogeneous} & \multicolumn{1}{c|}{$\times$} & \multicolumn{1}{c|}{} & \multicolumn{1}{c|}{IMPALA} & \multicolumn{1}{c|}{\checkmark} & \multicolumn{1}{c|}{\checkmark} & \multicolumn{1}{c|}{\checkmark} & \multicolumn{1}{c|}{$\times$} & \multicolumn{1}{c|}{Practical} \\ \cline{1-8} \cline{10-15}

\multicolumn{1}{|c|}{Zhou et al. \cite{zhou2023cost}} & \multicolumn{1}{c|}{Single} & \multicolumn{1}{c|}{Independent} & \multicolumn{1}{c|}{\Circle} & \multicolumn{1}{c|}{Homogeneous} & \multicolumn{1}{c|}{Edge and Cloud} & \multicolumn{1}{c|}{Homogeneous} & \multicolumn{1}{c|}{$\times$} & \multicolumn{1}{c|}{} & \multicolumn{1}{c|}{A3C} & \multicolumn{1}{c|}{\checkmark} & \multicolumn{1}{c|}{$\times$} & \multicolumn{1}{c|}{\checkmark} & \multicolumn{1}{c|}{$\times$} & \multicolumn{1}{c|}{Simulation} \\ \cline{1-8} \cline{10-15}

\multicolumn{1}{|c|}{Zhang et al. \cite{zhang2024lsia3cs}} & \multicolumn{1}{c|}{Single} & \multicolumn{1}{c|}{Independent} & \multicolumn{1}{c|}{\LEFTcircle} & \multicolumn{1}{c|}{Heterogeneous} & \multicolumn{1}{c|}{Edge and Cloud} & \multicolumn{1}{c|}{Heterogeneous} & \multicolumn{1}{c|}{$\times$} & \multicolumn{1}{c|}{} & \multicolumn{1}{c|}{A3C} & \multicolumn{1}{c|}{\checkmark} & \multicolumn{1}{c|}{$\times$} & \multicolumn{1}{c|}{$\times$} & \multicolumn{1}{c|}{$\times$} & \multicolumn{1}{c|}{Simulation} \\ \cline{1-8} \cline{10-15}

\multicolumn{1}{|c|}{Ju et al. \cite{ju2023noma}} & \multicolumn{1}{c|}{Single} & \multicolumn{1}{c|}{Independent} & \multicolumn{1}{c|}{\LEFTcircle} & \multicolumn{1}{c|}{Heterogeneous} & \multicolumn{1}{c|}{Edge} & \multicolumn{1}{c|}{Heterogeneous} & \multicolumn{1}{c|}{$\times$} & \multicolumn{1}{c|}{} & \multicolumn{1}{c|}{A3C} & \multicolumn{1}{c|}{$\times$} & \multicolumn{1}{c|}{\checkmark} & \multicolumn{1}{c|}{$\times$} & \multicolumn{1}{c|}{$\times$} & \multicolumn{1}{c|}{Simulation} \\ \cline{1-8} \cline{10-15}

\multicolumn{1}{|c|}{Liu et al. \cite{liu2023asynchronous}} & \multicolumn{1}{c|}{Single} & \multicolumn{1}{c|}{Independent} & \multicolumn{1}{c|}{\LEFTcircle} & \multicolumn{1}{c|}{Homogeneous} & \multicolumn{1}{c|}{Edge} & \multicolumn{1}{c|}{Homogeneous} & \multicolumn{1}{c|}{$\times$} & \multicolumn{1}{c|}{} & \multicolumn{1}{c|}{A3C} & \multicolumn{1}{c|}{\checkmark} & \multicolumn{1}{c|}{$\times$} & \multicolumn{1}{c|}{$\times$} & \multicolumn{1}{c|}{$\times$} & \multicolumn{1}{c|}{Simulation} \\ \cline{1-8} \cline{10-15}

\multicolumn{1}{|c|}{Shen et al. \cite{shen2024asynchronous}} & \multicolumn{1}{c|}{Multiple} & \multicolumn{1}{c|}{Dependent} & \multicolumn{1}{c|}{\LEFTcircle} & \multicolumn{1}{c|}{Heterogeneous} & \multicolumn{1}{c|}{Edge and Cloud} & \multicolumn{1}{c|}{Heterogeneous} & \multicolumn{1}{c|}{$\times$} & \multicolumn{1}{c|}{} & \multicolumn{1}{c|}{\begin{tabular}[c]{@{}c@{}}PPO + \\Fed Learning\end{tabular}} & \multicolumn{1}{c|}{\checkmark} & \multicolumn{1}{c|}{\checkmark} & \multicolumn{1}{c|}{\checkmark} & \multicolumn{1}{c|}{$\times$} & \multicolumn{1}{c|}{Simulation} \\ \cline{1-8} \cline{10-15}

\multicolumn{1}{|c|}{\textbf{KD-AFRL}} & \multicolumn{1}{c|}{Multiple} & \multicolumn{1}{c|}{Dependent} & \multicolumn{1}{c|}{\CIRCLE} & \multicolumn{1}{c|}{Heterogeneous} & \multicolumn{1}{c|}{Edge and Cloud} & \multicolumn{1}{c|}{Heterogeneous} & \multicolumn{1}{c|}{\checkmark} & \multicolumn{1}{c|}{} & \multicolumn{1}{c|}{\begin{tabular}[c]{@{}c@{}}Actor-Critic + \\Fed Learning + KD\end{tabular}} & \multicolumn{1}{c|}{\checkmark} & \multicolumn{1}{c|}{\checkmark} & \multicolumn{1}{c|}{\checkmark} & \multicolumn{1}{c|}{\checkmark} & \multicolumn{1}{c|}{Practical} \\ \hline

\multicolumn{15}{l}{\CIRCLE: Real IoT Application and Deployment, \LEFTcircle: Simulated IoT Application, \Circle: Random} 
\end{tabular}%
}
\end{table*}
\renewcommand{\arraystretch}{1}

\section{System Model and Problem Formulation}
\label{system_model}
This section first describes the topology of the Cloud-Edge-IoT multi-domain computing architecture. Next, we tackle the scheduling of IoT applications by formulating it as an optimization problem, aiming at
reducing application completion time, system energy consumption, and weighted cost. 

\subsection{System Model}
\begin{figure}[h]
\centering
\includegraphics[width=\linewidth]{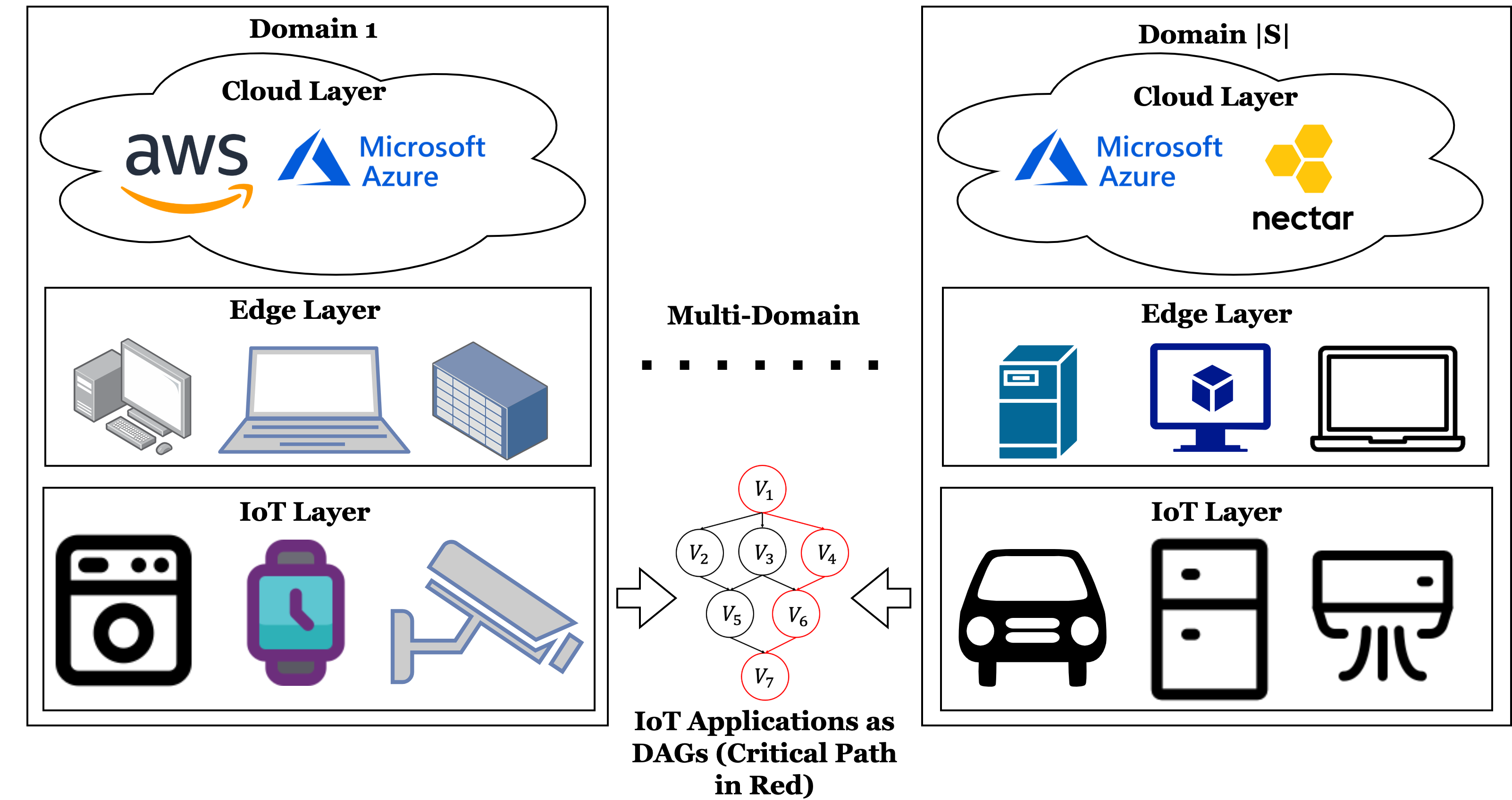}
\caption{The hierarchical Cloud-Edge-IoT computing architecture with distributed scheduling domains.}
\label{fig:ovw}
\end{figure}

Fig.~\ref{fig:ovw} depicts a hierarchical Cloud-Edge-IoT computing architecture with distributed scheduling domains. Our system comprises a heterogeneous computing infrastructure spanning cloud servers, edge nodes, and IoT devices, collectively forming a continuous computing environment.

The computing infrastructure consists of $|\mathcal{N}|$ servers, defined as $\mathcal{N} = \{\mathcal{N}_k|1 \leq k \leq |\mathcal{N}|\}$. To account for server heterogeneity, each server $\mathcal{N}_k$ is characterized by specific resource capabilities, including available CPU frequency $Freq(\mathcal{N}_k)$ measured in MHz and available memory $Ram(\mathcal{N}_k)$ measured in GB. The network connectivity between servers is defined by propagation time $\mathcal{P}_{\mathcal{N}_l, \mathcal{N}_k}$ (ms) and data transmission rate $\mathcal{B}_{\mathcal{N}_l, \mathcal{N}_k}$ (b/s).

A distinguishing feature of our system model is the distribution of scheduling responsibility across multiple domains. We define a set of schedulers $\mathcal{S} = \{\mathcal{S}_m|1 \leq m \leq |\mathcal{S}|\}$, where each scheduler $\mathcal{S}_m$ operates in a distinct environment with unique characteristics. Each scheduler $\mathcal{S}_m$ is responsible for managing a subset of computational resources $\mathcal{N}_m \subseteq \mathcal{N}$ and handling workloads within its domain. These domains may be geographically distributed, administratively separated, or functionally specialized, each exhibiting unique environmental dynamics that influence scheduling decisions.

Within each distributed domain, we consider a set of IoT applications $\mathcal{A} = \{\mathcal{A}_i|1 \leq i \leq |\mathcal{A}|\}$ that require execution across the available resources. Each application $\mathcal{A}_i$ consists of multiple interdependent tasks denoted as $\mathcal{A}_i = \{\mathcal{A}^j_i|1 \leq j \leq |\mathcal{A}_i|\}$. As illustrated in Fig.~\ref{fig:ovw}, each application is modeled as a DAG, where vertices $\mathcal{V}_j = \mathcal{A}^j_i$ represent individual tasks, and edges $\mathcal{E}_{j,k}$ represent data dependencies between tasks $\mathcal{V}_j$ and $\mathcal{V}_k$, indicating that successor tasks can only begin execution after their predecessors complete. The critical path, denoted as $CP(\mathcal{A}_i)$ and highlighted in red, represents the path with the highest cumulative cost from entry to exit tasks.

\subsection{Problem Formulation}
\label{problem formulation}
With multiple scheduling domains managing different subsets of resources, the scheduling process is distributed across various schedulers. For each task $\mathcal{A}^j_i$, we define its scheduling configuration as a tuple:
\begin{equation}
\mathcal{C}^j_{i} = (\mathcal{N}_k, \mathcal{S}_m), \quad k \in \{1, \dots, |\mathcal{N}|\}, \quad m \in \{1, \dots, |\mathcal{S}|\},
\end{equation}
where $\mathcal{N}_k$ denotes the server assigned to execute the task, and $\mathcal{S}_m$ represents the scheduler responsible for making this assignment. This configuration must satisfy the domain constraint: $\mathcal{N}_k \in \mathcal{N}_m$, ensuring that schedulers only allocate resources within their authority.

The scheduling configuration $\mathcal{C}_{i}$ for the application $\mathcal{A}_i$ encompasses all task-level configurations and is defined as:
\begin{equation}
\mathcal{C}_{i} = \{\mathcal{C}^j_{i} | 1 \leq j \leq |\mathcal{A}_i|\},
\end{equation}
where $|\mathcal{A}_i|$ represents the total number of tasks in application $\mathcal{A}_i$.

The execution model for applications preserves the dependency constraints represented in the DAG structure. Each task cannot begin execution until all its predecessor tasks complete. We use $PR(\mathcal{A}^j_i)$ to denote the set of predecessor tasks of task $\mathcal{A}^j_i$ and use $CP(\mathcal{A}^j_i)$ to indicate whether task $\mathcal{A}^j_i$ is located on the critical path of the application $\mathcal{A}_i$. 

\subsubsection{Application Completion Time Model}
Given the scheduling configuration $\mathcal{C}^j_{i} = (\mathcal{N}_k, \mathcal{S}_m)$ for task $\mathcal{A}^j_i$, we define the task completion time model $TCT(\mathcal{C}^j_{i})$ comprising two primary components: the communication latency model $T^{cl}(\mathcal{C}^j_{i})$ and the processing duration model $T^{pd}(\mathcal{C}^j_{i})$:
\begin{equation}
TCT(\mathcal{C}^j_{i}) = T^{cl}(\mathcal{C}^j_{i}) + T^{pd}(\mathcal{C}^j_{i}). 
\end{equation}

The communication latency model $T^{cl}(\mathcal{C}^j_{i})$ represents the maximum time required for data dependencies to be satisfied before task execution:
\begin{equation}
T^{cl}(\mathcal{C}^j_{i}) = \max_{\mathcal{A}^k_{i}\in PR(\mathcal{A}^j_i)}\;\;T_{\mathcal{C}^k_{i}, \mathcal{C}^j_{i}}^{cl},
\end{equation}
where $T_{\mathcal{C}^k_{i}, \mathcal{C}^j_{i}}^{cl}$ indicates the time needed to transfer data from the server executing predecessor task $\mathcal{A}^k_{i}$ to the server executing task $\mathcal{A}^j_i$. This time depends on both the data transfer time $T_{\mathcal{C}^k_{i}, \mathcal{C}^j_{i}}^{dt}$ and the network propagation time $\mathcal{P}_{\mathcal{N}_l, \mathcal{N}_k}$ between the respective servers:
\begin{equation} 
T_{\mathcal{C}^k_{i}, \mathcal{C}^j_{i}}^{cl} = 
\begin{cases}
T_{\mathcal{C}^k_{i}, \mathcal{C}^j_{i}}^{dt} + \mathcal{P}_{\mathcal{N}_l, \mathcal{N}_k}& \text{if servers differ},\\
0& \text{if same server},
\end{cases}
\end{equation}
where $\mathcal{N}_l$ and $\mathcal{N}_k$ are the servers in configurations $\mathcal{C}^k_{i}$ and $\mathcal{C}^j_{i}$ respectively. The data transfer time $T_{\mathcal{C}^k_{i}, \mathcal{C}^j_{i}}^{dt}$ is calculated as:
\begin{equation}
T_{\mathcal{C}^k_{i}, \mathcal{C}^j_{i}}^{dt} = \frac{DV_{\mathcal{C}^k_{i}, \mathcal{C}^j_{i}}(\mathcal{A}^j_i)}{\mathcal{B}_{\mathcal{N}_l, \mathcal{N}_k}},
\end{equation}
where $DV_{\mathcal{C}^k_{i}, \mathcal{C}^j_{i}}(\mathcal{A}^j_i)$ denotes the data volume for task $\mathcal{A}^j_i$ transmitted from the server in configuration $\mathcal{C}^k_{i}$ to the server in configuration $\mathcal{C}^j_{i}$, and $\mathcal{B}_{\mathcal{N}_l, \mathcal{N}_k}$ represents the data transmission rate between these servers.

The processing duration model $T^{pd}(\mathcal{C}^j_{i})$ defines the time required to execute task $\mathcal{A}^j_i$ on the assigned server and is calculated as:
\begin{equation}\label{eq:etm}
T^{pd}(\mathcal{C}^j_{i}) = \frac{CC(\mathcal{A}^j_i)}{Freq(\mathcal{N}_k)}, 
\end{equation}
where $CC(\mathcal{A}^j_i)$ denotes the computational complexity (in required CPU cycles) for executing task $\mathcal{A}^j_i$ and $Freq(\mathcal{N}_k)$ represents the CPU frequency of the assigned server $\mathcal{N}_k$ in configuration $\mathcal{C}^j_{i}$.

The completion time $CT(\mathcal{C}_{i})$ for the entire application $\mathcal{A}_i$ is expressed as:
\begin{equation}\label{eq:tts}
CT(\mathcal{C}_{i}) = \sum_{j=1}^{|\mathcal{A}_i|}(TCT(\mathcal{C}^j_{i}) \times CP(\mathcal{A}^j_i)), 
\end{equation}
where $CP(\mathcal{A}^j_i)$ is the critical path indicator: $1$ if task $\mathcal{A}^j_i$ is on the critical path of application $\mathcal{A}_i$, and $0$ otherwise.

For each scheduler $\mathcal{S}_m$, let $\mathcal{A}_{\mathcal{S}_m} = \{\mathcal{A}_i | \mathcal{A}_i \text{ is assigned to } \mathcal{S}_m\}$ denote the set of applications assigned to that scheduler. The global optimization objective is to minimize the sum of application completion times across all scheduling domains:
\begin{equation}
\min_{\mathcal{C}} \sum_{m=1}^{|\mathcal{S}|}\sum_{\mathcal{A}_i \in \mathcal{A}_{\mathcal{S}_m}} CT(\mathcal{C}_{i}),
\end{equation}
where $\mathcal{C}$ represents the collective scheduling decisions across all domains. 

\subsubsection{System Energy Consumption Model}
The energy consumption function $E(\cdot)$ characterizes the total energy required to execute applications across heterogeneous computing resources in a distributed environment. For a task $\mathcal{A}^j_i$ with scheduling configuration $\mathcal{C}^j_{i} = (\mathcal{N}_k, \mathcal{S}_m)$, the total energy consumption $E(\mathcal{C}^j_{i})$ consists of two components: the processing energy $E^{pd}(\mathcal{C}^j_{i})$ consumed during task execution, and the communication energy $E^{cm}(\mathcal{C}^j_{i})$ consumed when transmitting data to successor tasks:
\begin{equation}
E(\mathcal{C}^j_{i}) = E^{pd}(\mathcal{C}^j_{i}) + (E^{cm}(\mathcal{C}^j_{i}) \times ED(\mathcal{A}^j_i)),
\end{equation}
where $ED(\mathcal{A}^j_i)$ is a binary indicator: $0$ if $\mathcal{A}^j_i$ is a terminal task with no successors, and $1$ otherwise.

The processing energy model $E^{pd}(\mathcal{C}^j_{i})$ quantifies the energy consumed by the server when executing the task:
\begin{equation}
E^{pd}(\mathcal{C}^j_{i}) = T^{pd}(\mathcal{C}^j_{i}) \times P^{pd}(\mathcal{N}_k), 
\end{equation}
where $T^{pd}(\mathcal{C}^j_{i})$ is the processing duration obtained from the completion time model, and $P^{pd}(\mathcal{N}_k)$ represents the power consumption rate of server $\mathcal{N}_k$ during computation.

The communication energy model $E^{cm}(\mathcal{C}^j_{i})$ accounts for the energy expended when transmitting output data to the servers hosting successor tasks:
\begin{equation}\label{eq:tre}
E^{cm}(\mathcal{C}^j_{i}) = \sum\limits_{\mathcal{A}^l_{i} \in SU(\mathcal{A}^j_{i})}\frac{DV_{\mathcal{C}^j_{i}, \mathcal{C}^l_{i}}(\mathcal{A}^j_i)}{\mathcal{B}_{\mathcal{N}_k, \mathcal{N}_q}} \times P^{cm}(\mathcal{N}_k) \times \delta(\mathcal{N}_k, \mathcal{N}_q),
\end{equation}
where:
\begin{itemize}
\item $SU(\mathcal{A}^j_{i})$ denotes the set of successor tasks of task $\mathcal{A}^j_i$
\item $\mathcal{C}^l_{i} = (\mathcal{N}_q, \mathcal{S}_n)$ is the scheduling configuration of a successor task $\mathcal{A}^l_{i}$
\item $DV_{\mathcal{C}^j_{i}, \mathcal{C}^l_{i}}(\mathcal{A}^j_i)$ represents the volume of data transmitted
\item $\mathcal{B}_{\mathcal{N}_k, \mathcal{N}_q}$ is the data transmission rate between servers
\item $P^{cm}(\mathcal{N}_k)$ denotes the power consumption rate of server $\mathcal{N}_k$ during data transmission
\item $\delta(\mathcal{N}_k, \mathcal{N}_q)$ is a binary indicator: $0$ if $\mathcal{N}_k$ and $\mathcal{N}_q$ are the same server, and $1$ otherwise
\end{itemize}

The transmission power $P^{cm}(\mathcal{N}_k)$ can be modeled as a constant value for each server type or as a dynamic parameter that varies based on network conditions and transmission load.

The total energy consumption $E(\mathcal{C}_{i})$ for executing application $\mathcal{A}_i$ is calculated by summing the energy consumption of all constituent tasks:
\begin{equation}\label{eq:ets}
E(\mathcal{C}_{i}) = \sum_{j=1}^{|\mathcal{A}_i|}E(\mathcal{C}^j_{i}).
\end{equation}

For each scheduler $\mathcal{S}_m$, the global energy optimization objective is to minimize the sum of application energy consumption across all scheduling domains:
\begin{equation}
\min_{\mathcal{C}} \sum_{m=1}^{|\mathcal{S}|}\sum_{\mathcal{A}_i \in \mathcal{A}_{\mathcal{S}_m}} E(\mathcal{C}_{i}),
\end{equation}
where $\mathcal{C}$ represents the collective scheduling decisions across all domains. 

\subsubsection{Weighted Cost Model}
To address the multi-objective nature of scheduling in distributed environments, we define a weighted cost model that balances application completion time and system energy consumption. For each application $\mathcal{A}_i$ with scheduling configuration $\mathcal{C}_i$, the weighted cost model $J(\mathcal{C}_i)$ is defined as:
\begin{equation}\label{eq:wt}
\small
J(\mathcal{C}_i) = \alpha_{cost} \times \frac{CT(\mathcal{C}_i) - CT^{min}}{CT^{max} - CT^{min}} + (1-\alpha_{cost}) \times \frac{E(\mathcal{C}_i) - E^{min}}{E^{max} - E^{min}},
\end{equation}
where $CT^{min}$ and $CT^{max}$ represent the minimum and maximum achievable completion times, $E^{min}$ and $E^{max}$ represent the minimum and maximum achievable energy consumption values, and $\alpha_{cost} \in [0,1]$ is the weight parameter that controls the trade-off between completion time and energy efficiency. Normalization is necessary because completion time and energy consumption typically have different scales and units.

From a system-wide perspective, the global weighted cost optimization objective is to minimize the sum of weighted costs across all scheduling domains:
\begin{equation}
\min_{\mathcal{C}} \sum_{m=1}^{|\mathcal{S}|}\sum_{\mathcal{A}_i \in \mathcal{A}_{\mathcal{S}_m}} J(\mathcal{C}_{i}).
\end{equation}

This optimization problem is subject to the following constraints:
{\footnotesize
\begin{align}
\label{eq:cons}
     \text{s.t.} \quad 
     & C1:\;|\{\mathcal{N}_k | (\mathcal{N}_k, \mathcal{S}_m) = \mathcal{C}^j_i\}| = 1,\;\forall \mathcal{C}^j_i \in \mathcal{C}_i \\
     & C2:\;DV_{\mathcal{C}^k_{i}, \mathcal{C}^j_{i}}(\mathcal{A}^j_i), \mathcal{B}_{\mathcal{N}_l, \mathcal{N}_k} > 0,\; \forall \mathcal{N}_l, \mathcal{N}_k \in \mathcal{N}, \notag\\
     & \qquad\;\;\forall \mathcal{A}^j_{i} \in \mathcal{A}_{i} \\
     & C3:\;Freq(\mathcal{N}_k), Ram(\mathcal{N}_k) > 0,\; \forall \mathcal{N}_k \in \mathcal{N}\\
     & C4:\;\sum_{\mathcal{A}_i \in \mathcal{A}_{\mathcal{S}_m}}\sum_{\mathcal{A}^j_i \in \mathcal{A}_i}Ram(\mathcal{A}^j_i) \times SO(\mathcal{A}^j_i, \mathcal{N}_k) \notag \\
     & \qquad < Ram(\mathcal{N}_k), \;\forall \mathcal{N}_k\in \mathcal{N}_m, \forall \mathcal{S}_m \in \mathcal{S} \\
     & C5:\;TCT(\mathcal{A}^k_i) \geq TCT(\mathcal{A}^j_i) + T^{cl}_{\mathcal{C}^j_{i}, \mathcal{C}^k_{i}},\forall \mathcal{A}^j_i \in PR(\mathcal{A}^k_i) \\
     & C6:\;0 \leq \alpha_{cost} \leq 1
\end{align}
}

Constraint $C1$ ensures that each task is assigned to exactly one server. $C2$ specifies that data volume and bandwidth must be positive for all task communications. $C3$ defines lower bounds for server resources (CPU frequency and RAM). $C4$ ensures that every server has sufficient RAM to process all tasks scheduled on it, where $SO(\mathcal{A}^j_i, \mathcal{N}_k)$ equals 1 if task $\mathcal{A}^j_i$ is scheduled on server $\mathcal{N}_k$ and 0 otherwise. $C5$ enforces precedence constraints, ensuring that a task can only start after its predecessors complete and the necessary data is transferred. Finally, $C6$ restricts the weight parameter to values between 0 and 1.

This optimization problem presents significant challenges due to its non-convex nature, time-varying constraints, and heterogeneous multi-domain environment. Traditional optimization methods and heuristic algorithms struggle with these complexities, particularly when adapting to dynamic resource availability and handling cross-domain interactions \cite{cui2024latency}. DRL offers a promising alternative by adapting to changing conditions without requiring complete system knowledge. This approach allows us to formulate scheduling as a sequential decision process that naturally balances immediate and long-term performance goals.

\subsection{Deep Reinforcement Learning Formulation}
To solve the complex multi-domain scheduling optimization problem, we formulate it as a DRL problem where each scheduler $\mathcal{S}_m$ learns an optimal scheduling policy through interactions with its environment. We model this as a Markov Decision Process (MDP) defined by the tuple $\langle \mathcal{S}, \mathcal{A}, \mathcal{P}, \mathcal{R}, \gamma \rangle$, where $\mathcal{S}$ represents the state space, $\mathcal{A}$ is the action space, $\mathcal{P}$ denotes the state transition probability function that determines how the system state evolves after executing an action, defined as $P(s_{t+1}|s_t,a_t) = Pr[S_{t+1} = s_{t+1}|S_t = s_t, A_t = a_t]$, $\mathcal{R}$ is the reward function, and $\gamma \in [0,1]$ is the discount factor.

\subsubsection{State Space}
In the multi-domain scheduling environment, the state observation for scheduler $\mathcal{S}_m$ at time step $t$ is defined as a triplet:
\begin{equation}
s_t^m = \{SR_t^m, AT_t^m, QT_t^m\},
\end{equation}
where:
\begin{itemize}
\item $SR_t^m = \{sr_1, sr_2, ..., sr_{|\mathcal{N}_m|}\}$ represents the current status of all servers in domain $m$, with each $sr_i$ including CPU utilization, CPU frequency, available memory, network load, server type identifier (cloud, edge, or IoT), and bandwidth.

\item $AT_t^m = \{at_1, at_2, ..., at_k\}$ describes attributes of the current task to be scheduled, including task ID, application ID, required CPU cycles, memory requirements, and data dependencies (predecessor tasks $PR(\mathcal{A}^j_i)$ and successor tasks $SU(\mathcal{A}^j_i)$).

\item $QT_t^m = \{qt_1, qt_2, ..., qt_q\}$ captures the task queue status, including waiting tasks, execution status of predecessor tasks, and scheduling locations of configured tasks.
\end{itemize}

\subsubsection{Action Space}
For each scheduler $\mathcal{S}_m$, the action at time step $t$ is defined as:
\begin{equation}
a_t^m = \{n_k | n_k \in \mathcal{N}_m\}.
\end{equation}

This action represents the decision to schedule the current task on server $n_k$ within domain $m$. The action space is discrete with cardinality equal to the number of available servers in the domain.

\subsubsection{Reward Function}
The reward function directly connects to our optimization objective, providing feedback on scheduling decision quality. For scheduler $\mathcal{S}_m$, the reward at time step $t$ is:
\begin{equation}
r_t^m = 
\begin{cases}
-J(\mathcal{C}_i) & \text{if task execution succeeds}, \\
\text{penalty} & \text{if task execution fails}.
\end{cases}
\end{equation}

Here, $J(\mathcal{C}_i)$ is the weighted cost model from Eq. \ref{eq:wt}, and the negative sign converts our minimization problem into a reward maximization problem. The penalty for failed executions encourages the agent to avoid decisions that lead to task failures.

\subsubsection{Policy and Objective}
The agent's policy function defines the probability of selecting action $a$ when observing state $s$:
\begin{equation}
\pi(a|s) = Pr[A_t=a|S_t=s].
\end{equation}

The ultimate goal of each scheduler $\mathcal{S}_m$ is to learn a policy $\pi_m$ that maximizes the expected cumulative discounted reward:
\begin{equation}
\pi_m^* = \arg\max_{\pi_m} \mathbb{E}\left[\sum_{t=0}^{\infty} \gamma^t r_t^m | \pi_m\right].
\end{equation}

This DRL formulation transforms our complex optimization problem into an iterative learning process. However, the multi-domain nature introduces several challenges. First, heterogeneity in resource characteristics and workload patterns leads to significantly different state distributions across domains. Second, the non-IID data across domains makes direct policy sharing ineffective. Third, domain-specific constraints and varying computational capabilities necessitate adaptable model architectures. In the following section, we introduce our Knowledge Distillation-empowered Adaptive Federated Reinforcement Learning (KD-AFRL) framework, which addresses these challenges while enabling collaborative learning across domains.

\section{KD-AFRL Framework}
\label{kd-afrl}
The KD-AFRL framework addresses challenges in multi-domain scheduling by combining federated learning with knowledge distillation techniques. 

The KD-AFRL framework operates across $M$ scheduling domains $\{\mathcal{S}_1, \mathcal{S}_2, ..., \mathcal{S}_M\}$, where each domain $\mathcal{S}_m$ possesses different computational resources $\mathcal{N}_m$, and workload characteristics. The framework addresses several key challenges in multi-domain reinforcement learning through three principal mechanisms:

\begin{itemize}
\item \textbf{Resource-Aware Hybrid Model Architecture Generation} addresses device heterogeneity by adapting DRL model complexity to match the computational capabilities of each domain's devices, enabling participation in federated learning regardless of resource constraints.

\item \textbf{Privacy-Preserving Environment-Clustered Federated Learning} enhances federated learning by identifying similar domains through differentially-private environmental features, addressing the non-IID challenges in distributed environments.

\item \textbf{Environment-Oriented Cross-Architecture Knowledge Distillation} complements the federated framework by enabling knowledge transfer between heterogeneous models based on environmental similarities, allowing resource-constrained devices to benefit from complex models without sharing raw data.
\end{itemize}

These mechanisms operate synergistically within an iterative learning paradigm, blending distributed experience collection with coordinated federated optimization. 

\subsection{Resource-Aware Hybrid Architecture Generation}
\begin{figure}[h]
\centering
\includegraphics[width=\linewidth]{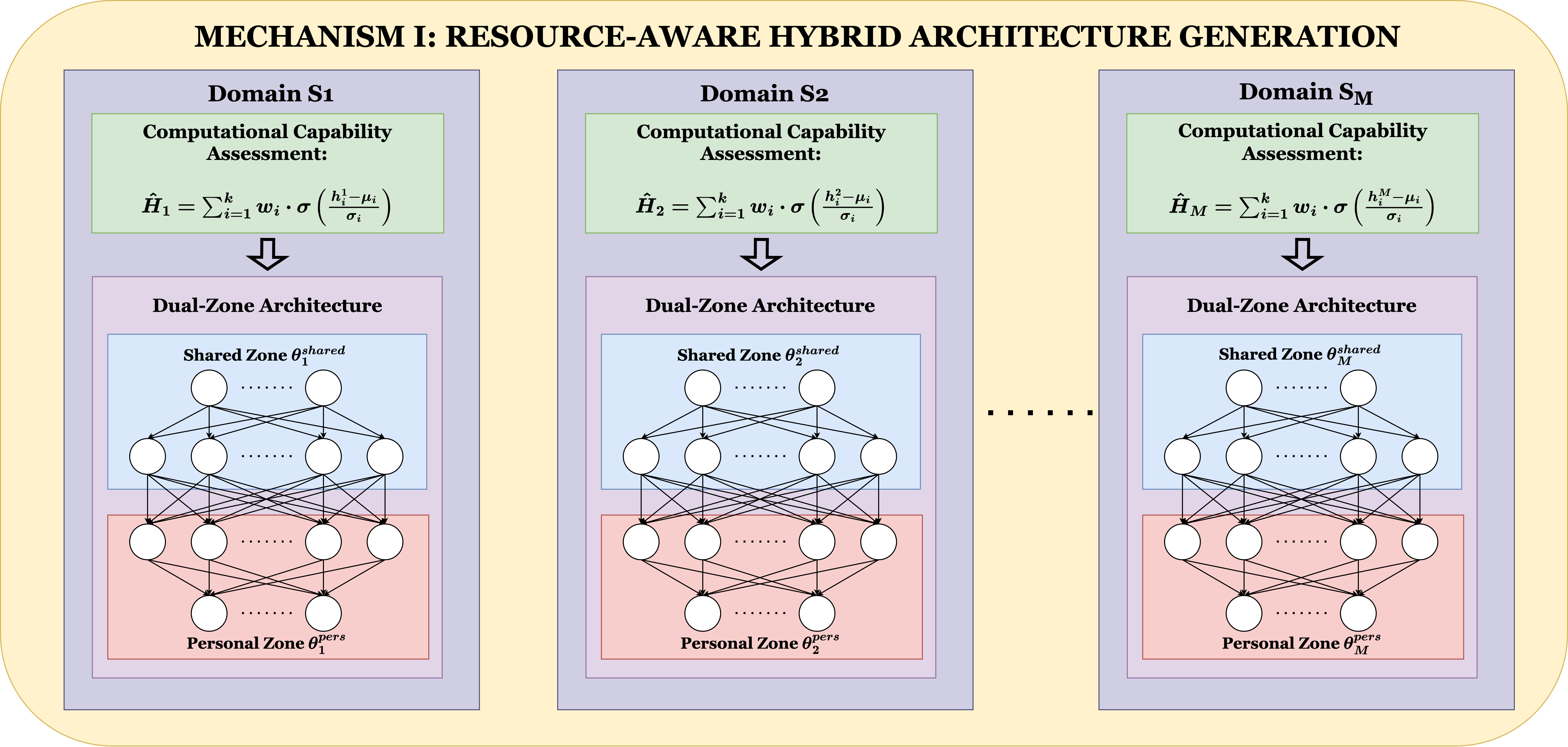}
\caption{Resource-aware hybrid architecture generation mechanism showing computational capability assessment and dual-zone architecture design across heterogeneous domains.}
\label{fig:mech1}
\end{figure}

To address device heterogeneity across domains while enabling collaborative learning, KD-AFRL incorporates a resource-aware model architecture generation mechanism that creates hybrid neural network structures combining standardized shared foundational zones with personalized adaptation zones, as shown in Fig.~\ref{fig:mech1}.

\subsubsection{Multi-Dimensional Computational Capability Assessment}
To enable precise and adaptive tailoring of model architectures, we quantify each scheduler's computational capability through a comprehensive resource profiling mechanism. For a scheduler in domain $\mathcal{S}_m$, its capability is represented by the vector $H_m \in \mathbb{R}^k$:

\begin{equation}
H_m = [h_1^m, h_2^m, ..., h_k^m]^T,
\end{equation}
where each dimension corresponds to a critical computational resource, such as CPU cores, CPU frequency (GHz), memory capacity (MB), and network bandwidth (Mbps).

To reflect real-time resource availability, we incorporate dynamic utilization effects:

\begin{equation}
h_i^m = \alpha_i \cdot h_i^{m,\text{static}} \cdot (1 - \beta_i \cdot \text{util}_i^m).
\end{equation}

Here, $\alpha_i$ is a weight coefficient for resource $i$, $h_i^{m,\text{static}}$ denotes the scheduler's inherent capability, $\beta_i$ is a sensitivity factor capturing the impact of utilization (set to 0 if the resource dimension does not have an associated utilization metric), and $\text{util}_i^m$ is the current utilization ratio. This formulation ensures that as resource utilization increases, the effective capacity diminishes proportionally, capturing transient resource constraints in dynamic environments.

We compute a scheduler's comprehensive capability score via a standardized non-linear aggregation:

\begin{equation}
\hat{H}_m = \sum_{i=1}^{k} w_i \cdot \sigma\left(\frac{h_i^m - \mu_i}{\sigma_i}\right),
\end{equation}
where $\sigma(\cdot)$ is the sigmoid function, $\mu_i$ and $\sigma_i$ are the mean and standard deviation of resource $i$ across all schedulers, and $w_i$ denotes its importance weight. This standardization ensures fair cross-resource comparison, while the sigmoid mapping provides robustness to outliers and maintains sensitivity across diverse resource scales.

\subsubsection{Dual-Zone Architecture Design}

To enable federated learning while accommodating resource heterogeneity, each domain's model is partitioned into a dual-zone structure with shared foundational zones (participating in federated aggregation) and personalized adaptation zones (retained locally):

\begin{equation}
\theta_m = \{\theta_m^{shared}, \theta_m^{pers}\}.
\end{equation}
The shared foundational zone $\theta_m^{shared}$ is selected from $K_{arch}$ predefined architecture types to ensure federated aggregation compatibility, while the personalized adaptation zone $\theta_m^{pers}$ is tailored to exploit remaining computational resources for domain-specific optimization.

\textbf{Shared Foundational Zone Selection}: Based on the capability score $\hat{H}_m$, domain $m$ is assigned to one of $K_{arch}$ predefined shared architecture types using a capability-based mapping:

\begin{equation}
\text{Type}_m = \min\left(\lceil \hat{H}_m \cdot K_{arch} \rceil, K_{arch}\right).
\end{equation}

This mapping function divides the normalized capability range [0,1] into $K_{arch}$ equal intervals, where each interval corresponds to a specific shared architecture type. The ceiling function $\lceil \cdot \rceil$ ensures integer type assignment, while the $\min(\cdot, K_{arch})$ operation prevents exceeding the maximum type number. For example, with $K_{arch}=3$ types and a capability score $\hat{H}_m = 0.7$, the assignment becomes $\text{Type}_m = \min(\lceil 0.7 \times 3 \rceil, 3) = \min(\lceil 2.1 \rceil, 3) = \min(3, 3) = 3$, assigning the domain to the most complex shared architecture type.

\textbf{Personalized Adaptation Zone Design}: The personalized adaptation zone leverages the domain's computational capability to provide domain-specific optimization. The zone's architecture scales proportionally with the capability score:

\begin{equation}
\theta_m^{depth,pers} = \lceil d_{\min} + \alpha_{arch} \cdot \hat{H}_m \cdot (d_{\max} - d_{\min}) \rceil,
\end{equation}

\begin{equation}
\theta_m^{width,pers} = \lceil w_{\min} + \alpha_{arch} \cdot \hat{H}_m \cdot (w_{\max} - w_{\min}) \rceil,
\end{equation}
where $\theta_m^{depth,pers}$ represents the number of layers in the personalized zone, $\theta_m^{width,pers}$ denotes the number of neurons per layer, $d_{\min}$, $d_{\max}$, $w_{\min}$, and $w_{\max}$ define the architectural parameter ranges, and $\alpha_{arch} \in [0,1]$ is a scaling factor that controls the personalized zone's complexity relative to the domain's total computational capacity.

The detailed procedure for resource-aware hybrid architecture generation is outlined in Algorithm \ref{ag:MAG}. The algorithm first assesses each domain's computational capability through multi-dimensional resource profiling, then assigns appropriate shared architecture types based on capability scores, and finally designs personalized adaptation zones scaled according to available computational resources. This ensures optimal resource utilization while maintaining federated learning compatibility.

\subsection{Privacy-Preserving Environment-Clustered Federated Learning}
\begin{figure}[t]
\centering
\includegraphics[width=\linewidth]{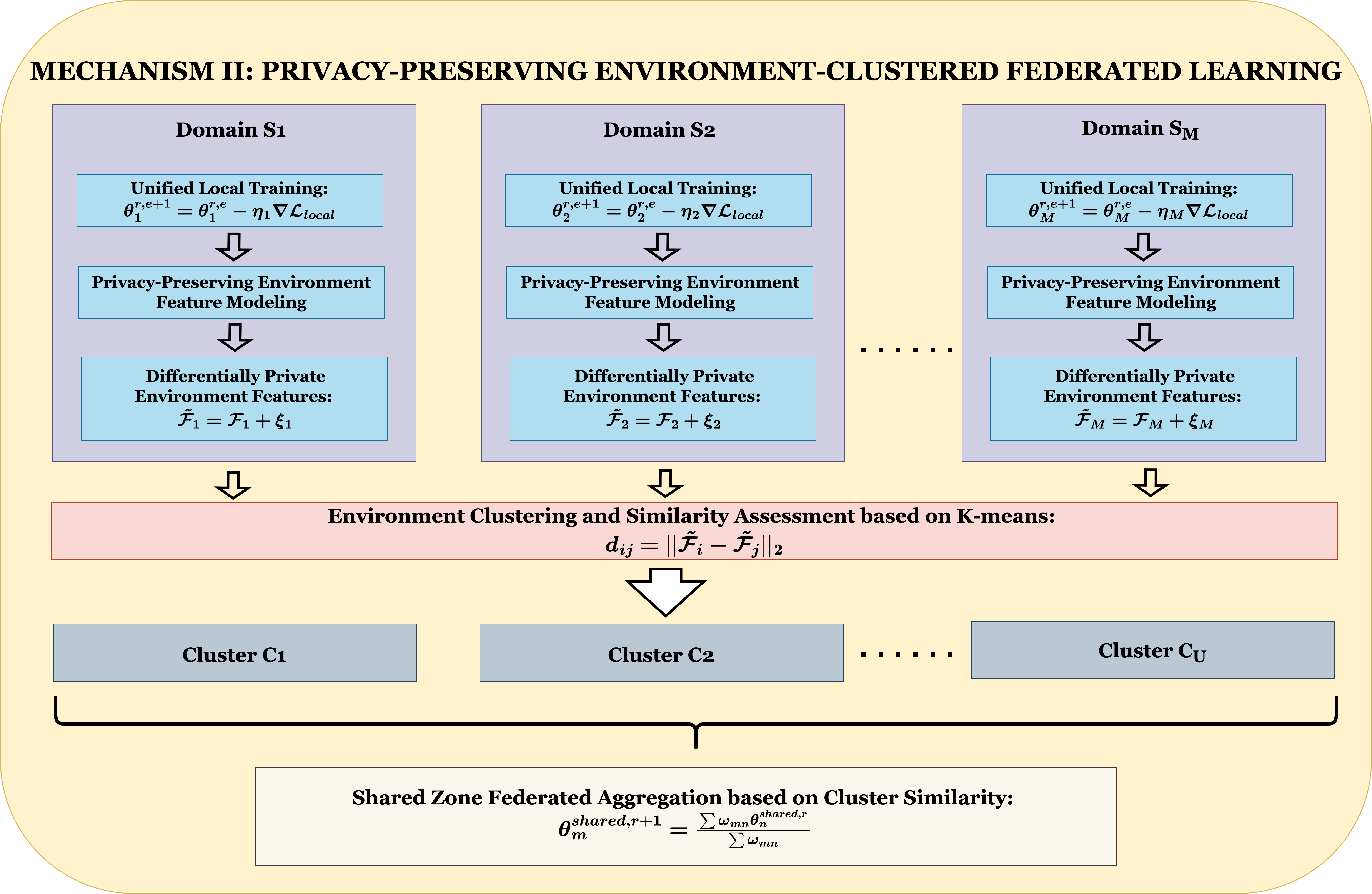}
\caption{Privacy-preserving environment-clustered federated learning mechanism illustrating unified local training, differential privacy protection, environment clustering, and shared zone aggregation.}
\label{fig:mech2}
\end{figure}

Traditional federated learning assumes that all participating domains share similar data distributions, which rarely holds in practice for multi-domain IoT scheduling environments. Different scheduling domains often exhibit distinct workload patterns, resource characteristics, and environmental dynamics, leading to significant performance degradation when directly applying conventional federated averaging. To address this challenge, KD-AFRL incorporates a privacy-preserving environment-clustered federated learning mechanism that identifies similar scheduling domains while protecting sensitive operational information, and coordinates the training of hybrid dual-zone architectures, as shown in Fig.~\ref{fig:mech2}.

\subsubsection{Privacy-Preserving Environment Feature Modeling}

To enable meaningful similarity assessment across scheduling domains, we extract comprehensive environment features that capture the operational characteristics of each domain. For scheduler $\mathcal{S}_m$, the environment feature vector $\mathcal{F}_m \in \mathbb{R}^d$ is constructed from two key dimensions:

\textbf{Resource Characteristics}: These features capture the computational landscape of domain $m$, including the mean and variance of CPU and memory utilization, inter-server communication bandwidth, and energy consumption (average power per computational unit).

\textbf{Workload Patterns}: These metrics characterize application workload dynamics, including the ratio of average task count to application count, standard deviation of completion times, average task arrival rate, and average dependency ratio in DAG structures.

To protect sensitive operational information while enabling collaborative learning, we implement a differential privacy mechanism by adding calibrated noise to the environment features:
\begin{equation}
\tilde{\mathcal{F}}_m = \mathcal{F}_m + \xi_m,
\end{equation}
where $\xi_m \sim \mathcal{L}(0, 1/\epsilon)$ is zero-mean Laplace noise calibrated to ensure $\epsilon$-differential privacy, with $\epsilon$ controlling the privacy level (smaller $\epsilon$ provides stronger privacy protection).

\begin{algorithm}[t]
\footnotesize
\caption{Resource-Aware Hybrid Architecture Generation}\label{ag:MAG}
\SetAlgoLined
\DontPrintSemicolon
\textbf{Input:} Resource capabilities $\{H_m\}_{m=1}^M$, architecture parameters $K_{arch}$, $d_{\min}$, $d_{\max}$, $w_{\min}$, $w_{\max}$, scaling factor $\alpha$\;
\textbf{Output:} Final hybrid architectures $\{\theta_m\}_{m=1}^M$ with type assignments\;
\textbf{Initialization:}\;
Initialize predefined shared architectures $\{\theta_k^{shared}\}_{k=1}^{K_{arch}}$ and parameter ranges\;
Compute resource statistics $\mu_i$ and $\sigma_i$ across all schedulers for each resource dimension\;
\textbf{Architecture Generation:}\;
\For{each scheduler $m = 1,2,...,M$}{
    Calculate normalized capability score $\hat{H}_m$\;
    Assign shared architecture type: $\text{Type}_m = \min(\lceil \hat{H}_m \cdot K_{arch} \rceil, K_{arch})$\;
    Select corresponding shared foundational zone: $\theta_m^{shared} = \theta_{\text{Type}_m}^{shared}$\;
    Design personalized adaptation zone using capability-based scaling:\;
    $\theta_m^{depth,pers} = \lceil d_{\min} + \alpha \cdot \hat{H}_m \cdot (d_{\max} - d_{\min}) \rceil$\;
    $\theta_m^{width,pers} = \lceil w_{\min} + \alpha \cdot \hat{H}_m \cdot (w_{\max} - w_{\min}) \rceil$\;
    Set final architecture: $\theta_m = \{\theta_m^{shared}, \theta_m^{pers}\}$\;
}
\Return{Final hybrid architectures $\{\theta_m\}_{m=1}^M$ with type assignments}
\end{algorithm}

\subsubsection{Environment Clustering and Similarity Assessment}

Given the privacy-preserving environment features $\{\tilde{\mathcal{F}}_1, \tilde{\mathcal{F}}_2, ..., \tilde{\mathcal{F}}_M\}$ from all participating domains, the central server performs K-means clustering to identify groups of similar scheduling environments. The similarity between domains is measured using Euclidean distance:
\begin{equation}
d_{ij} = ||\tilde{\mathcal{F}}_i - \tilde{\mathcal{F}}_j||_2.
\end{equation}
This partitions the domains into $U$ clusters $\{C_1, C_2, ..., C_U\}$. Despite the added Laplace noise for privacy protection, the clustering remains effective as the noise has zero mean and gets averaged out during the iterative process.

To handle the dynamic nature of scheduling environments, we implement a drift-based adaptation mechanism. We periodically compute the drift for each domain:
\begin{equation}
\text{Drift}_m^{(r)} = ||\mathcal{F}_m^{(r)} - \mathcal{F}_m^{(r-\Delta r)}||_2.
\end{equation}
When the maximum drift across all domains exceeds the threshold:
\begin{equation}
\max_m \text{Drift}_m^{(r)} > \tau_{drift},
\end{equation}
we trigger re-clustering to ensure that domains with substantially changed environments are appropriately reassigned to better-matching clusters.

\subsubsection{Dual-Zone Coordinated Learning Strategy}
Based on the clustering results and the hybrid dual-zone architectures, we design a coordinated learning strategy that combines unified local training with shared zone federated aggregation. During local training, both shared and personalized zones are optimized using domain-specific data, while only shared zones participate in cross-domain knowledge sharing through federated aggregation.

\textbf{Unified Local Training}: During local training phases, each domain performs gradient updates on the complete dual-zone model using local data. This ensures that both shared and personalized zones are optimized jointly while maintaining model coherence:
\begin{equation}
\theta_m^{r,e+1} = \theta_m^{r,e} - \eta_{m} \nabla_{\theta_m} \mathcal{L}_{local}(\theta_m^{r,e}),
\end{equation}
where $\theta_m^{r,e}$ represents the complete model parameters for domain $m$ at round $r$ and local epoch $e$, $\eta_{m}$ is the learning rate for domain $m$, and $\mathcal{L}_{local}$ is the local loss function computed on domain $m$'s data using the complete dual-zone model.

\textbf{Shared Zone Federated Aggregation}: After local training, only the shared foundational zones $\theta_m^{shared}$ participate in federated aggregation, grouped by their architecture types. Domains with the same shared architecture type (determined by $\text{Type}_m$) can directly aggregate their learned parameters through weighted averaging. 

For domain $m$ belonging to environmental cluster $C_j$, the similarity weight with respect to any other domain $n$ is computed as:
\begin{equation}
\small  
\omega_{mn} = 
\begin{cases}
\exp\left(-\frac{d_{mn}^2}{2\sigma_j^2}\right) & \text{if } \text{Type}_n = \text{Type}_m \text{ and } n \in C_j, \\
\beta_{fed} \cdot \exp\left(-\frac{d_{mn}^2}{2\sigma_{global}^2}\right) & \text{if } \text{Type}_n = \text{Type}_m \text{ and } n \notin C_j, \\
0 & \text{if } \text{Type}_n \neq \text{Type}_m,
\end{cases}
\end{equation}
where $\sigma_j^2$ is the intra-cluster variance within cluster $C_j$, $\sigma_{global}^2$ is the global variance across all domains, and $\beta_{fed} \in [0,1]$ is a cross-cluster damping factor that reduces the influence of domains from different environmental clusters. The variance terms $\sigma_j^2$ and $\sigma_{global}^2$ serve as adaptive scaling factors that normalize distances relative to their respective typical scales (intra-cluster and global), ensuring appropriate weight calibration across different cluster densities and domain distributions.

\begin{algorithm}[t]
\footnotesize
\caption{Privacy-Preserving Environment-Clustered Federated Learning}\label{ag:PPECFL}
\SetAlgoLined
\DontPrintSemicolon
\textbf{Input:} Environment features $\{\mathcal{F}_m\}_{m=1}^M$, hybrid architectures $\{\theta_m\}_{m=1}^M$, privacy budget $\epsilon$, number of clusters $U$, drift threshold $\tau_{drift}$\;
\textbf{Output:} Cluster assignments $\{C_u\}_{u=1}^U$, trained models $\{\theta_m^{shared}, \theta_m^{pers}\}_{m=1}^M$\;
\textbf{Initial Environment Clustering:}\;
\ForPar{each domain $m = 1,2,...,M$}{
    Extract environment features $\mathcal{F}_m^{(0)}$\;
    Generate privacy-preserving features: $\tilde{\mathcal{F}}_m^{(0)} = \mathcal{F}_m^{(0)} + \xi_m$ where $\xi_m \sim \mathcal{L}(0, 1/\epsilon)$\;
    Send $\tilde{\mathcal{F}}_m^{(0)}$ to central server\;
}
Perform K-means clustering on $\{\tilde{\mathcal{F}}_m^{(0)}\}$ to obtain initial clusters $\{C_u\}_{u=1}^U$\;
\textbf{Dynamic Dual-Zone Federated Learning:}\;
\For{communication round $r = 1,2,...,R$}{
    \ForPar{each domain $m$}{
        \For{local epoch $e = 1,2,...,E_{local}$}{
            Update complete model: $\theta_m \leftarrow \theta_m - \eta_m \nabla_{\theta_m} \mathcal{L}_{local}$\;
        }
    }
    \If{$r \bmod T_{fed} = 0$}{
        \textbf{Environment Drift Detection and Re-clustering:}\;
        \ForPar{each domain $m$}{
            Extract current environment features $\mathcal{F}_m^{(r)}$\;
            Generate privacy-preserving features: $\tilde{\mathcal{F}}_m^{(r)} = \mathcal{F}_m^{(r)} + \xi_m$\;
            Send $\tilde{\mathcal{F}}_m^{(r)}$ to central server\;
        }
        \For{each domain $m$}{
            Compute drift: $\text{Drift}_m^{(r)} = ||\tilde{\mathcal{F}}_m^{(r)} - \tilde{\mathcal{F}}_m^{(r-T_{fed})}||_2$\;
        }
        \If{$\max_m \text{Drift}_m^{(r)} > \tau_{drift}$}{
            Update clusters: Perform K-means clustering on $\{\tilde{\mathcal{F}}_m^{(r)}\}$ to obtain updated $\{C_u\}_{u=1}^U$\;
        }
        \textbf{Federated Aggregation:}\;
        \ForPar{each domain $m$}{
            Send locally-trained $\theta_m^{shared}$ to central server\;
        }
        \For{each domain $m$}{
            Compute similarity weights $\omega_{mn}$ using current clusters $\{C_u\}$\;
            Aggregate shared parameters: $\theta_m^{shared,r+1} = \frac{\sum_{n=1}^M \omega_{mn} \theta_n^{shared,r}}{\sum_{n=1}^M \omega_{mn}}$\;
        }
        Send aggregated $\theta_m^{shared,r+1}$ back to each domain $m$\;
    }
}
\Return{Final cluster assignments $\{C_u\}_{u=1}^U$, trained models $\{\theta_m^{shared}, \theta_m^{pers}\}_{m=1}^M$}
\end{algorithm}

This weight assignment ensures that: (1) only domains with identical shared architecture types can participate in parameter aggregation; (2) domains within the same environmental cluster have higher influence on each other; and (3) cross-cluster knowledge sharing is maintained but appropriately dampened to prevent interference from dissimilar environments. 

The aggregated shared zone parameters $\theta_m^{shared,r+1}$ for domain $m$ at round $r+1$ are computed as:
\begin{equation}
\theta_m^{shared,r+1} = \frac{\sum_{n=1}^M \omega_{mn} \cdot \theta_n^{shared,r}}{\sum_{n=1}^M \omega_{mn}}.
\end{equation}

The detailed procedure for the privacy-preserving environment-clustered federated learning is outlined in Algorithm \ref{ag:PPECFL}. The algorithm first performs environment clustering with differential privacy protection, then alternates between local training (where both zones learn from local data) and federated aggregation for shared zones (with similarity-weighted parameter averaging based on environmental compatibility). This approach enables collaborative learning among similar domains while preserving domain-specific adaptations and protecting sensitive operational information.

\subsection{Environment-Oriented Cross-Architecture Knowledge Distillation}
\begin{figure}[h]
\centering
\includegraphics[width=\linewidth]{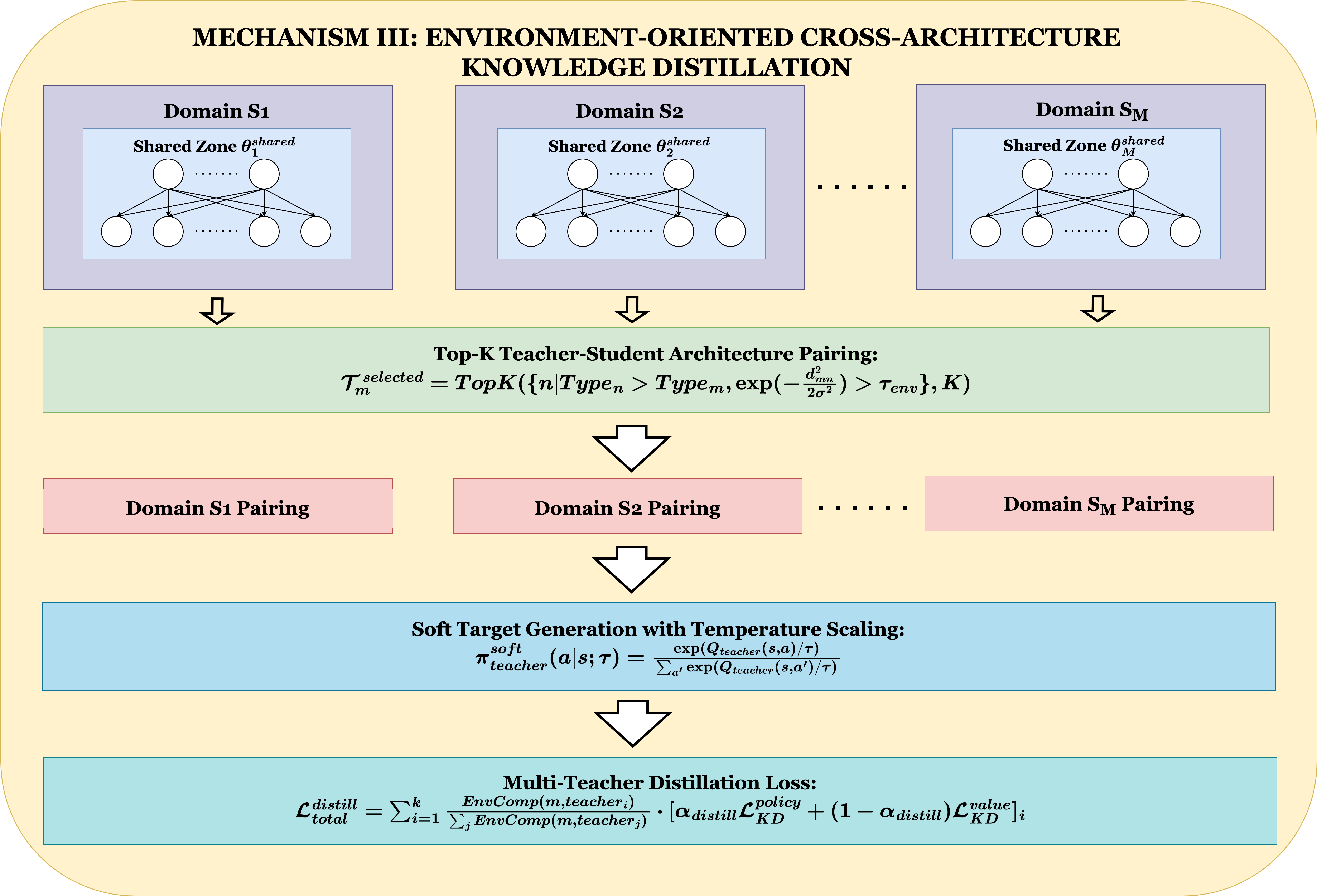}
\caption{Environment-oriented cross-architecture knowledge distillation mechanism demonstrating Top-K teacher-student pairing and multi-teacher distillation process.}
\label{fig:mech3}
\end{figure}

While privacy-preserving environmental clustering federated learning enables collaboration among domains with identical shared architecture types, domains with different architecture types cannot directly participate in parameter aggregation due to structural incompatibility, preventing resource-constrained domains from benefiting from knowledge gained by more capable domains. To address this challenge, KD-AFRL introduces an environment-oriented cross-architecture knowledge distillation mechanism that enables knowledge transfer between domains with heterogeneous model architectures while considering environmental similarities, as shown in Fig.~\ref{fig:mech3}. This mechanism enables effective knowledge transfer between models with heterogeneous architectures, allowing small models on resource-constrained devices to achieve near-comparable performance to larger models.

\subsubsection{Top-K Teacher-Student Architecture Pairing Strategy}

The knowledge distillation process operates through teacher-student relationships, where domains with more complex architectures (teachers) transfer knowledge to domains with simpler architectures (students). Given $K_{arch}$ predefined shared architecture types ranked by complexity, we establish a Top-K teacher-student pairing strategy.

For a domain $\mathcal{S}_m$ with architecture type $\text{Type}_m$, its potential teacher set $\mathcal{T}_m$ includes all domains with higher complexity architecture types in the same or similar environmental clusters:

\begin{equation}
\mathcal{T}_m = \{n \mid \text{Type}_n > \text{Type}_m, \text{EnvComp}(m,n) > \tau_{env}\},
\end{equation}
where $\text{EnvComp}(m,n)$ denotes the environmental compatibility between domains $m$ and $n$:
\begin{equation}
\text{EnvComp}(m,n) = \exp\left(-\frac{d_{mn}^2}{2\sigma_{global}^2}\right),
\end{equation}
and $\tau_{env} \in [0,1]$ is a threshold parameter controlling the minimum environmental similarity required for teacher-student pairing.

The server automatically establishes distillation pairing relationships based on model architecture complexity. Based on complexity ranking, each student model selects the Top-K models with higher complexity as its teacher set. The teacher selection strategy is defined as:

 \begin{equation}
\mathcal{T}_m^{selected} = \text{TopK}\left(\mathcal{T}_m, K\right).
\end{equation}

\subsubsection{Soft Target Knowledge Distillation Mechanism}

The core challenge of cross-architecture knowledge distillation lies in transferring strategic knowledge between models with different structural configurations. We design a soft target-based dual-level knowledge distillation method that achieves cross-architecture knowledge transfer through the transmission of policy distributions and value estimates.

\textbf{Temperature-Regulated Soft Target Generation}: To capture the complete decision knowledge of teacher models, we employ temperature regulation techniques to generate soft targets. For a given state $s_t$, the teacher model's raw Q-value output is $Q_{teacher}(s_t, a)$. By introducing a temperature parameter $\tau_{temp}$, we convert these Q-values into a softened probability distribution:

\begin{equation}
\pi_{teacher}^{soft}(a|s_t; \tau_{temp}) = \frac{\exp(Q_{teacher}(s_t, a)/\tau_{temp})}{\sum_{a' \in \mathcal{A}} \exp(Q_{teacher}(s_t, a')/\tau_{temp})}.
\end{equation}

The temperature parameter $\tau_{temp}$ controls the smoothness of the distribution and the granularity of knowledge transfer: when $\tau_{temp} > 1$, the probability distribution becomes more smooth, reducing the dominance of optimal actions and enabling the student model to learn the teacher's relative preferences for suboptimal actions, which helps transfer richer decision knowledge; when $\tau_{temp} = 1$, it degrades to the standard softmax distribution.

\textbf{Policy Knowledge Distillation}: The student model learns the policy distribution by minimizing the Kullback-Leibler divergence with the teacher's soft targets:
\begin{equation}
\begin{aligned}
\mathcal{L}_{KD}^{policy} &= \mathbb{E}_{s_t \sim \mathcal{D}_{student}} \Big[ D_{KL}\big(\pi_{teacher}^{soft}(a|s_t; \tau_{temp}) \\
&\qquad \parallel \pi_{student}(a|s_t; \tau_{temp})\big) \Big] \times \tau_{temp}^2,
\end{aligned}
\end{equation}
where the $\tau_{temp}^2$ term is a gradient compensation factor to maintain consistent loss scaling across different temperature values.

\textbf{Value Function Knowledge Transfer}: In addition to policy distributions, value functions contain important estimation information about long-term rewards. The value function distillation loss is defined as:

\begin{equation}
\mathcal{L}_{KD}^{value} = \mathbb{E}_{s_t \sim \mathcal{D}_{student}} \left[ \left(V_{teacher}(s_t) - V_{student}(s_t)\right)^2 \right].
\end{equation}

\textbf{Adaptive Distillation Loss Balancing}: To balance the contributions of policy distillation and value function distillation, the complete distillation loss function for student domains is:

\begin{equation}
\mathcal{L}_{distill} = \alpha_{distill} \cdot \mathcal{L}_{KD}^{policy} + (1-\alpha_{distill}) \cdot \mathcal{L}_{KD}^{value},
\end{equation}
where $\alpha_{distill} \in [0,1]$ is a balancing parameter.

\subsubsection{Environmental Compatibility-Oriented Multi-Teacher Distillation Strategy}

Considering the environmental differences between different domains, we design an environmental compatibility-based multi-teacher distillation strategy.

\textbf{Teacher Weight Assignment}: Considering the environmental differences between different domains, we design an environmental compatibility-based teacher weight assignment strategy. For student model $m$, the weight of its $i$-th teacher is based on environmental similarity:

\begin{equation}
w_i = \frac{\text{EnvComp}(m, teacher_i)}{\sum_{j=1}^{k} \text{EnvComp}(m, teacher_j)}.
\end{equation}

This design ensures that teachers with similar environments receive higher weights, as knowledge from teacher models in similar environments is more easily transferred to student models.

\textbf{Training Episode Allocation}: The distillation training episodes for different teachers are adaptively allocated based on their environmental compatibility:

\begin{equation}
E_i = E_{base} \cdot \left(1 + \beta_{distill} \cdot \text{EnvComp}(m, teacher_i)\right),
\end{equation}
where $E_{base}$ is the base training episodes and $\beta_{distill}$ is an adjustment parameter. This design ensures that teachers with similar environments receive more training episodes.

\textbf{Multi-Teacher Loss Aggregation}: Based on environmental compatibility weights, the total distillation loss is:

\begin{equation}
\mathcal{L}_{total}^{distill} = \sum_{i=1}^{k} w_i \cdot \mathcal{L}_{distill}^{i},
\end{equation}
where $\mathcal{L}_{distill}^{i}$ is the distillation loss from the $i$-th teacher.

The complete cross-architecture knowledge distillation process is outlined in Algorithm \ref{ag:CAKD}. The algorithm first establishes teacher-student pairings based on architecture complexity and environmental compatibility, then performs multi-teacher distillation where each student learns from multiple teachers with weights and training episodes allocated according to environmental similarity. This ensures that students prioritize learning from teachers in similar environments while still benefiting from diverse architectural knowledge.

\begin{algorithm}[htb!]
\footnotesize
\caption{Environment-Oriented Cross-Architecture Knowledge Distillation}\label{ag:CAKD}
\SetAlgoLined
\DontPrintSemicolon
\textbf{Input:} Domains with architectures $\{\theta_m\}_{m=1}^M$, architecture types $\{\text{Type}_m\}_{m=1}^M$, environmental compatibility threshold $\tau_{env}$, number of teachers $K$, base temperature $\tau_{temp}$, adjustment parameter $\beta_{distill}$, balancing parameter $\alpha_{distill}$\;
\textbf{Output:} Enhanced models $\{\theta_m\}_{m=1}^M$ through knowledge distillation\;
\textbf{Initialize Top-K Teacher-Student Pairing:}\;
Sort by model architecture complexity in descending order: $\{m_1, m_2, ..., m_M\}$\;
\For{each student model $m_j$, $j = 2$ to $M$}{
    $\mathcal{T}_{m_j} = \{m_i | i < j, \text{EnvComp}(m_j, m_i) > \tau_{env}\}$\;
    $\mathcal{T}_{m_j}^{selected} = \text{TopK}(\mathcal{T}_{m_j}, K)$\;
}
\textbf{Multi-Teacher Knowledge Distillation Training:}\;
\For{each distillation round $r = 1,2,...,R_{KD}$}{
    \ForPar{each student model $m_j$}{
        \For{each teacher $m_i \in \mathcal{T}_{m_j}^{selected}$, $i = 1$ to $k$}{
            Set training episodes: $E_i = E_{base} \cdot (1 + \beta_{distill} \cdot \text{EnvComp}(m_j, m_i))$\;
            Compute teacher weight: $w_i = \frac{\text{EnvComp}(m_j, m_i)}{\sum_{l=1}^{k} \text{EnvComp}(m_j, m_l)}$\;
            \For{training episode $e = 1$ to $E_i$}{
                Obtain teacher targets: $\pi_{teacher}^{soft}(a|s; \tau_{temp})$, $V_{teacher}(s)$\;
                Compute policy distillation loss: $\mathcal{L}_{KD}^{policy,i} = D_{KL}(\pi_{teacher}^{soft} \parallel \pi_{student}) \times \tau_{temp}^2$\;
                Compute value distillation loss: $\mathcal{L}_{KD}^{value,i} = ||V_{teacher} - V_{student}||^2$\;
                Compute single teacher loss: $\mathcal{L}_{distill}^{i} = \alpha_{distill} \cdot \mathcal{L}_{KD}^{policy,i} + (1-\alpha_{distill}) \cdot \mathcal{L}_{KD}^{value,i}$\;
                Accumulate weighted loss: $\mathcal{L}_{total}^{distill} += w_i \cdot \mathcal{L}_{distill}^{i}$\;
            }
        }
        Update student model: $\theta_{student} \leftarrow \theta_{student} - \eta_{student} \nabla_{\theta_{student}} \mathcal{L}_{total}^{distill}$\;
    }
}
\Return{Enhanced models $\{\theta_m\}_{m=1}^M$}
\end{algorithm}

\section{Performance Evaluation}
\label{evaluation}
This section introduces the experimental setup, hyperparameter tuning configurations, and comprehensive experiments to evaluate KD-AFRL performance.

\subsection{Experiment Setup}
This subsection describes the distributed multi-domain experimental environment, IoT application workloads, and baseline techniques.

\subsubsection{Practical Experiment Environment}
\begin{figure}[t]
\centering
\includegraphics[width=\linewidth]{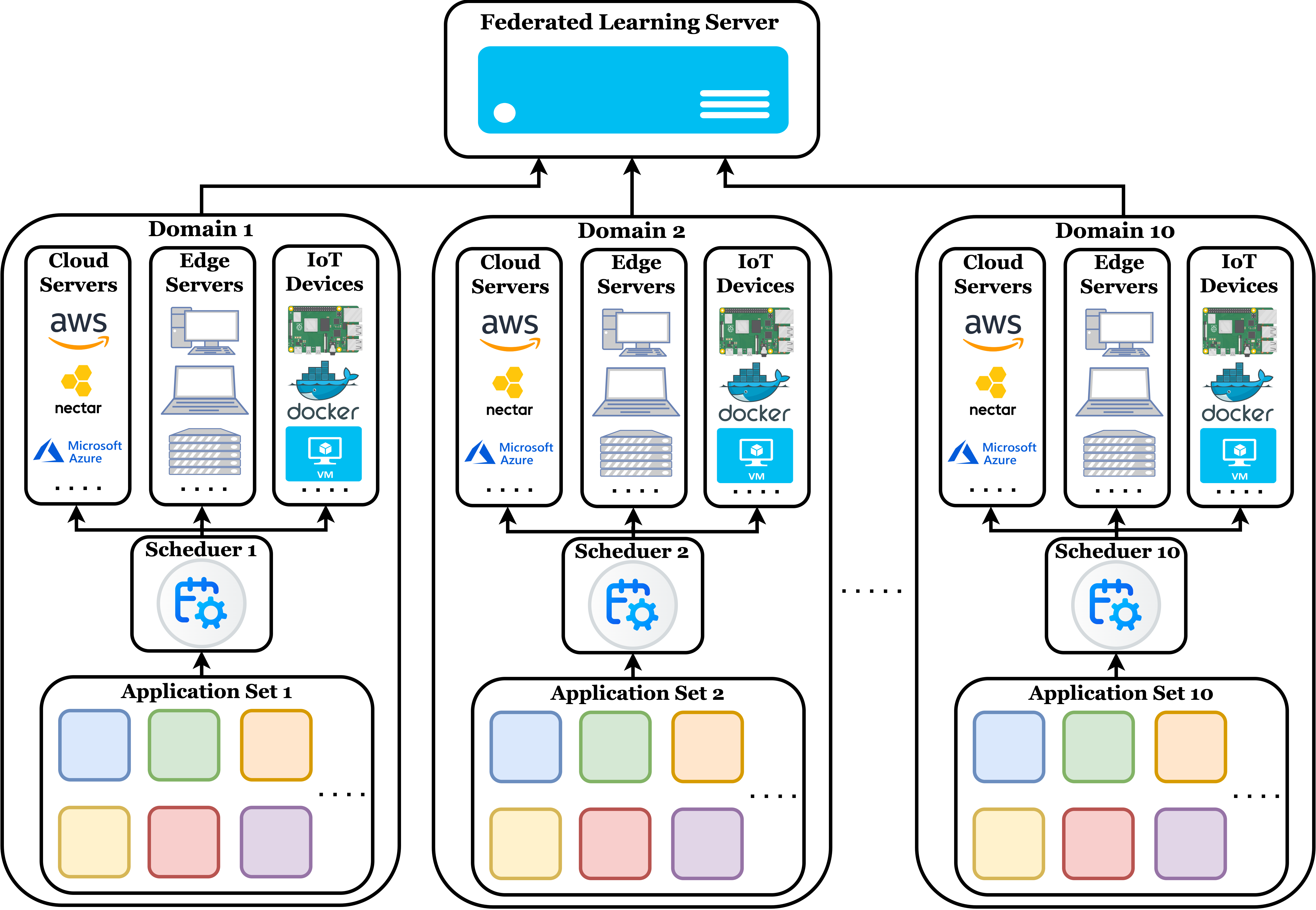}
\caption{KD-AFRL experimental environment}
\label{fig:exp}
\end{figure}

To evaluate the effectiveness of KD-AFRL in realistic heterogeneous multi-domain environments, we establish a distributed experimental infrastructure consisting of 10 independent scheduling domains across different geographical locations and infrastructure providers, as shown in Fig. \ref{fig:exp}. Each domain operates as an autonomous entity with an independent scheduler responsible for resource management and workload scheduling, containing various combinations of cloud servers, edge servers, and IoT devices to form a heterogeneous Cloud-Edge-IoT computing environment.

At the Cloud layer, we deploy instances from different cloud service providers, including Nectar Cloud instances (AMD EPYC processors, configurations ranging from 2 cores @2.0GHz 8GB RAM to 32 cores @2.0GHz 128GB RAM), AWS Cloud instances (Intel Xeon processors, configurations ranging from 1 core @2.4GHz 1GB RAM to 16 cores @2.5GHz 64GB RAM), and Microsoft Azure Cloud instances (Intel Xeon processors, configurations ranging from 1 core @2.3GHz 1GB RAM to 24 cores @2.4GHz 96GB RAM). At the Edge layer, we configure various edge computing devices, including devices based on M1 Pro processors (8 cores, 16GB RAM), devices equipped with Intel Core i7 processors (8 cores @2.3GHz, 16GB RAM), Intel Core i9 processors (8 cores @2.5GHz, 32GB RAM), and Intel Core i5 processors (6 cores @2.8GHz, 8GB RAM) with different configurations. At the IoT layer, we deploy Raspberry Pi devices (Pi OS, Broadcom BCM2837 quad-core @1.2GHz, 1GB RAM), virtual machines and Docker containers equipped with webcams and IP cameras. 

The network connections exhibit realistic latency and bandwidth variations, reflecting real-world deployment scenarios. The latency between IoT devices and edge nodes ranges from 1-6ms with bandwidth of 10-25 MB/s. Network characteristics between IoT devices and cloud nodes vary by cloud service provider: latency with Nectar cloud servers ranges from 6-12ms with bandwidth of 14-20MB/s; latency with AWS cloud servers ranges from 15-25ms with bandwidth of 15-22MB/s; latency with Microsoft Azure cloud servers ranges from 7-15ms with bandwidth of 15-21MB/s. Communication latency between edge nodes and cloud nodes ranges from 6-25ms with bandwidth of 15-22MB/s. For energy monitoring, we use the \texttt{eco2AI} \cite{budennyy2022eco2ai} to implement accurate real-time power measurement. In Equation~\ref{eq:wt}, we set the weight parameter $\alpha_{cost}$ to 0.5 to balance response time and energy consumption optimization objectives.

\subsubsection{IoT Application Workloads}
To rigorously evaluate KD-AFRL under diverse computational conditions, we construct a heterogeneous IoT workload suite that reflects real-world deployments:
\begin{itemize}
  \item \textit{AudioAmplitudeMonitor}: real-time amplitude tracking implemented with \texttt{librosa} \cite{mcfee2015librosa}; workload scaled by analysis-window length.
  \item \textit{TextSentimentAnalysis}: sentiment inference on text using \texttt{TextBlob} \cite{loria2018textblob} and \texttt{NLTK} \cite{bird2006nltk}; workload scaled by text-block size.
  \item \textit{SpeechRecognition}: speech-to-text inference combining \texttt{torchaudio} \cite{yang2022torchaudio} and a lightweight Transformer decoder; workload scaled by audio-chunk length.
  \item \textit{DataCompressionService}: lossless compression with \texttt{zlib} \cite{gailly1995zlib}/\texttt{gzip} \cite{rfc1952}; workload scaled by file size and compression level.
  \item \textit{ImageProcessor}: batch image filtering and resizing using \texttt{PIL} \cite{clark2015pillow} and \texttt{OpenCV} \cite{bradski2000opencv}; workload scaled by batch size.
  \item \textit{HealthTracker}: activity recognition via \texttt{TensorFlow~Lite} \cite{abadi2016tensorflow, tflite2017} pose estimation; workload scaled by sampling rate and model capacity.
  \item \textit{FaceDetection}: face detection on streaming video with \texttt{OpenCV} \cite{bradski2000opencv} and \texttt{MediaPipe} \cite{lugaresi2019mediapipe}; workload scaled by frame resolution and detection frequency.
  \item \textit{ColorTracking}: HSV-based colour tracking for AR overlays using \texttt{OpenCV} \cite{bradski2000opencv}; workload scaled by frame rate and tracking-window size.
  \item \textit{FaceAndEyeDetection}: cascaded face-and-eye detection with landmark refinement using \texttt{OpenCV} \cite{bradski2000opencv} and \texttt{dlib} \cite{king2009dlib}; workload scaled by input resolution and cascade depth.
  \item \textit{VideoOCR}: video text spotting (detection + recognition) with \texttt{EasyOCR} \cite{easyocr2020} and \texttt{OpenCV} \cite{bradski2000opencv}; workload scaled by clip length and model precision.
\end{itemize}

By systematically sweeping each application's dominant parameters (e.g., window length, batch size, model precision), we generate diverse IoT applications that span the full spectrum of resource footprints—covering I/O-bound, CPU-bound, memory-intensive, GPU-accelerated, and network-constrained characteristics—thereby exercising KD-AFRL across realistic deployment scenarios.

\subsubsection{Baseline Techniques}
We implement KD-AFRL and all baseline techniques on the ReinFog platform \cite{wang2025reinfog} to ensure consistent experimental conditions. ReinFog is a modular platform for DRL-based resource management that supports both centralized and distributed techniques. To comprehensively evaluate the effectiveness of KD-AFRL, we adapt four representative state-of-the-art techniques to our multi-domain Cloud-Edge-IoT computing environment:
\begin{itemize}
\item \textbf{AFO}: The adapted version of the technique proposed in \cite{shen2024asynchronous}. This technique employs federated DRL for task offloading. We adapted it by modifying its architecture and resource allocation models to operate in our heterogeneous Cloud-Edge-IoT computing environments. Additionally, we revised its reward function to align with our optimization objectives.

\item \textbf{TF-DDRL}: The extended version of the technique proposed in \cite{wang2025tf}. This technique employs Transformer-enhanced distributed DRL based on IMPALA for IoT application scheduling in heterogeneous edge and cloud computing environments. We extended its architecture to support multi-domain scheduling scenarios. 

\item \textbf{ADTO}: The adapted version of the technique proposed in \cite{zhao2024asynchronous}. This technique employs A3C to solve the task offloading problem. We adapted its architecture to suit our multi-domain Cloud-Edge-IoT task scheduling problem. We also updated its reward function to align with our optimization objectives. Additionally, as A3C is commonly used in current literature (\cite{zhou2023cost, zhang2024lsia3cs, ju2023noma, liu2023asynchronous}), ADTO provides a representative benchmark for evaluating A3C-based solutions.

\item \textbf{DRLIS}: The extended version of the technique proposed in \cite{wang2024deep}. This technique employs a centralized DRL agent based on PPO for IoT application scheduling. We extended its reward function to align with our optimization objectives. As PPO is a policy gradient (PG) algorithm with superior training stability and sample efficiency \cite{yu2022surprising}, and given that PG algorithms are commonly used in current literature (\cite{zhu2025drl, fan2025vehicular, hsieh2023deep, zhao2023meson}), DRLIS provides a representative benchmark for evaluating PG-based solutions.
\end{itemize}

\begin{figure*}[h]
\begin{subfigure}{0.33\textwidth}
  \centering
  \includegraphics[width=\linewidth]{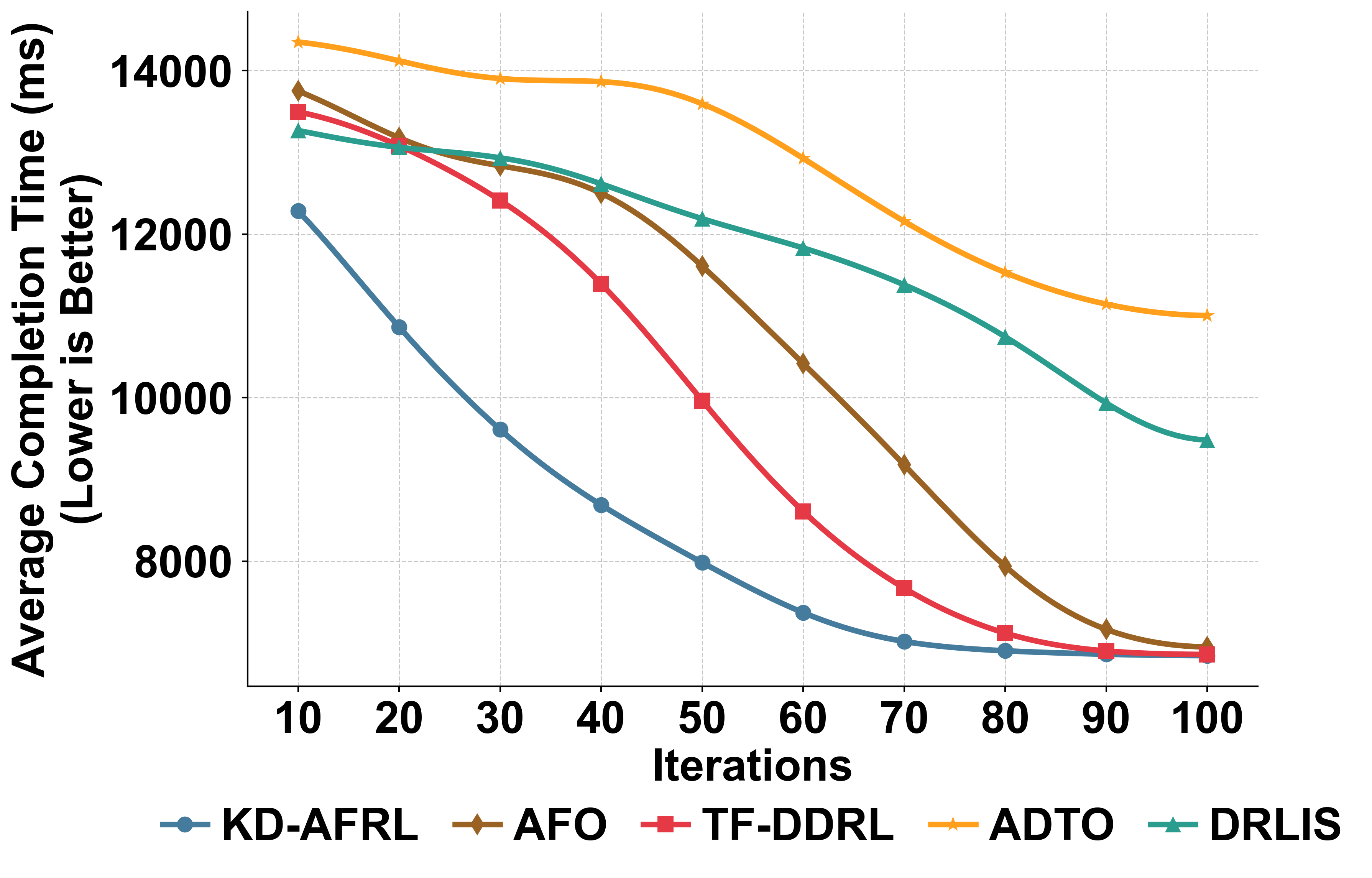}
  \caption{Average completion time}
  \label{fig:ctt}
\end{subfigure}%
\begin{subfigure}{0.33\textwidth}
  \centering
  \includegraphics[width=\linewidth]{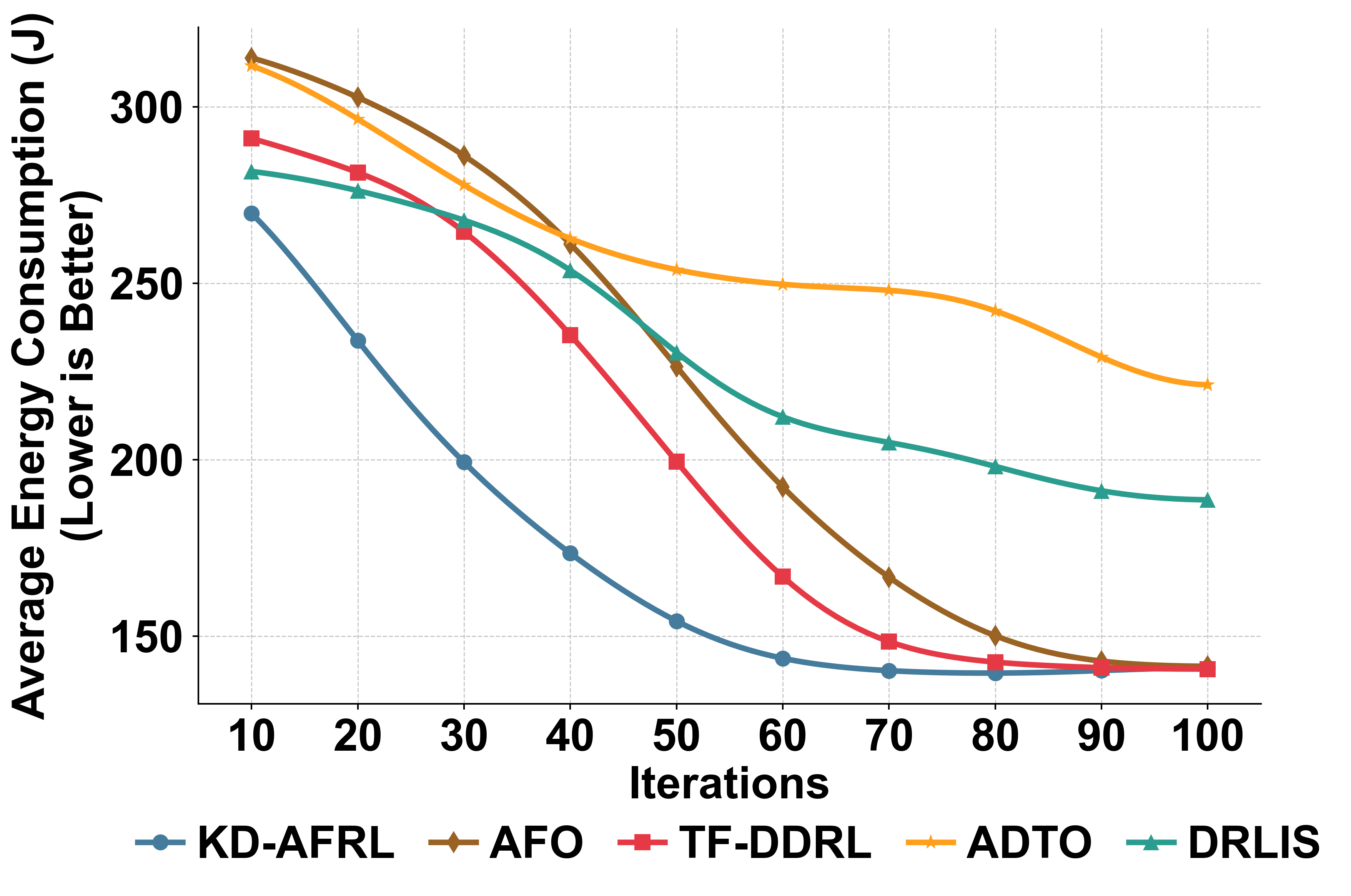}
  \caption{Average energy consumption}
  \label{fig:ect}
\end{subfigure}
\begin{subfigure}{0.33\textwidth}
  \centering
  \includegraphics[width=\linewidth]{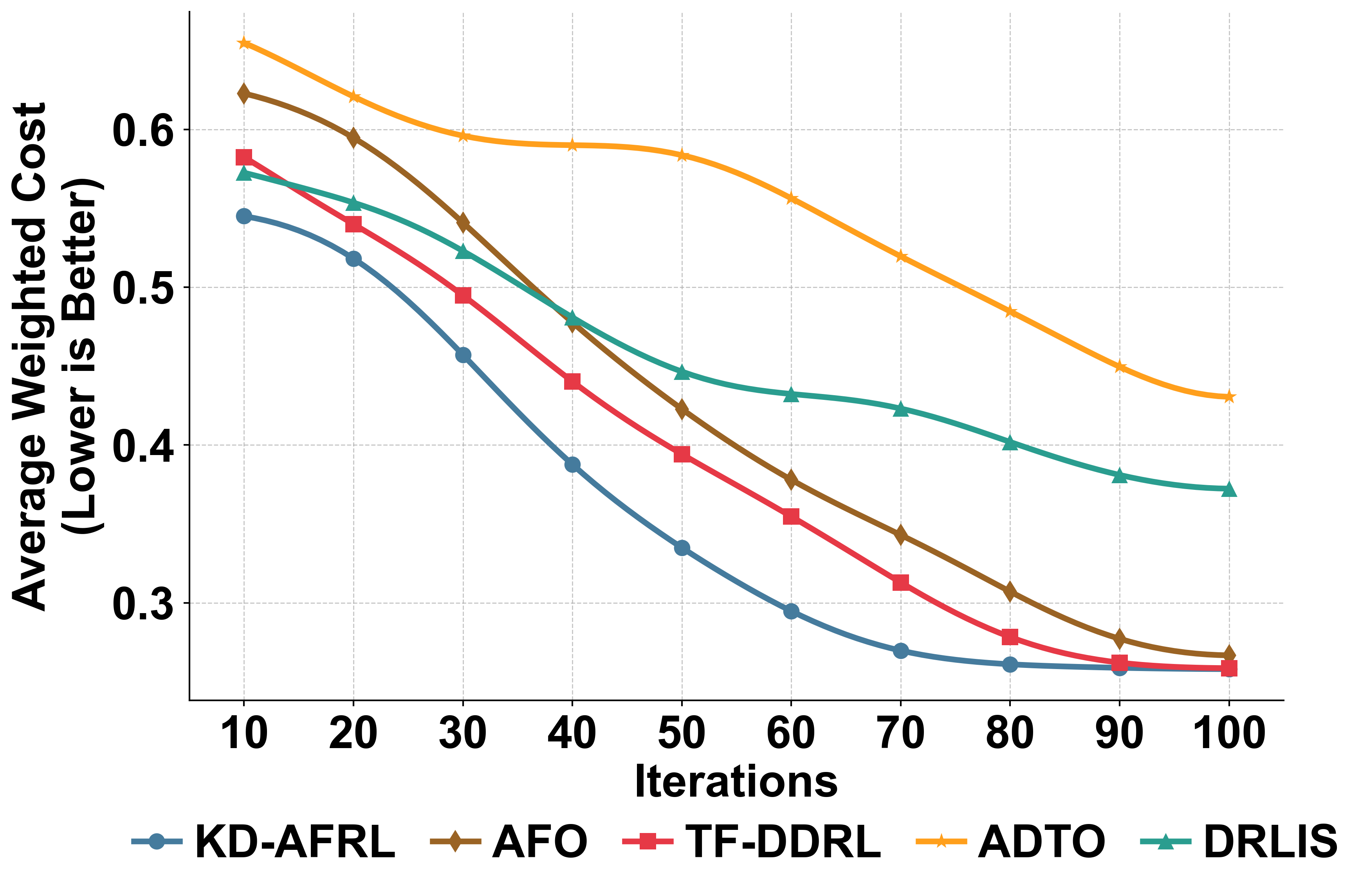}
  \caption{Average weighted cost}
  \label{fig:wct}
\end{subfigure}
\caption{Convergence performance analysis during training phase}
\label{fig:cpa1}
\end{figure*}
\begin{figure*}[h]
\begin{subfigure}{0.33\textwidth}
  \centering
  \includegraphics[width=\linewidth]{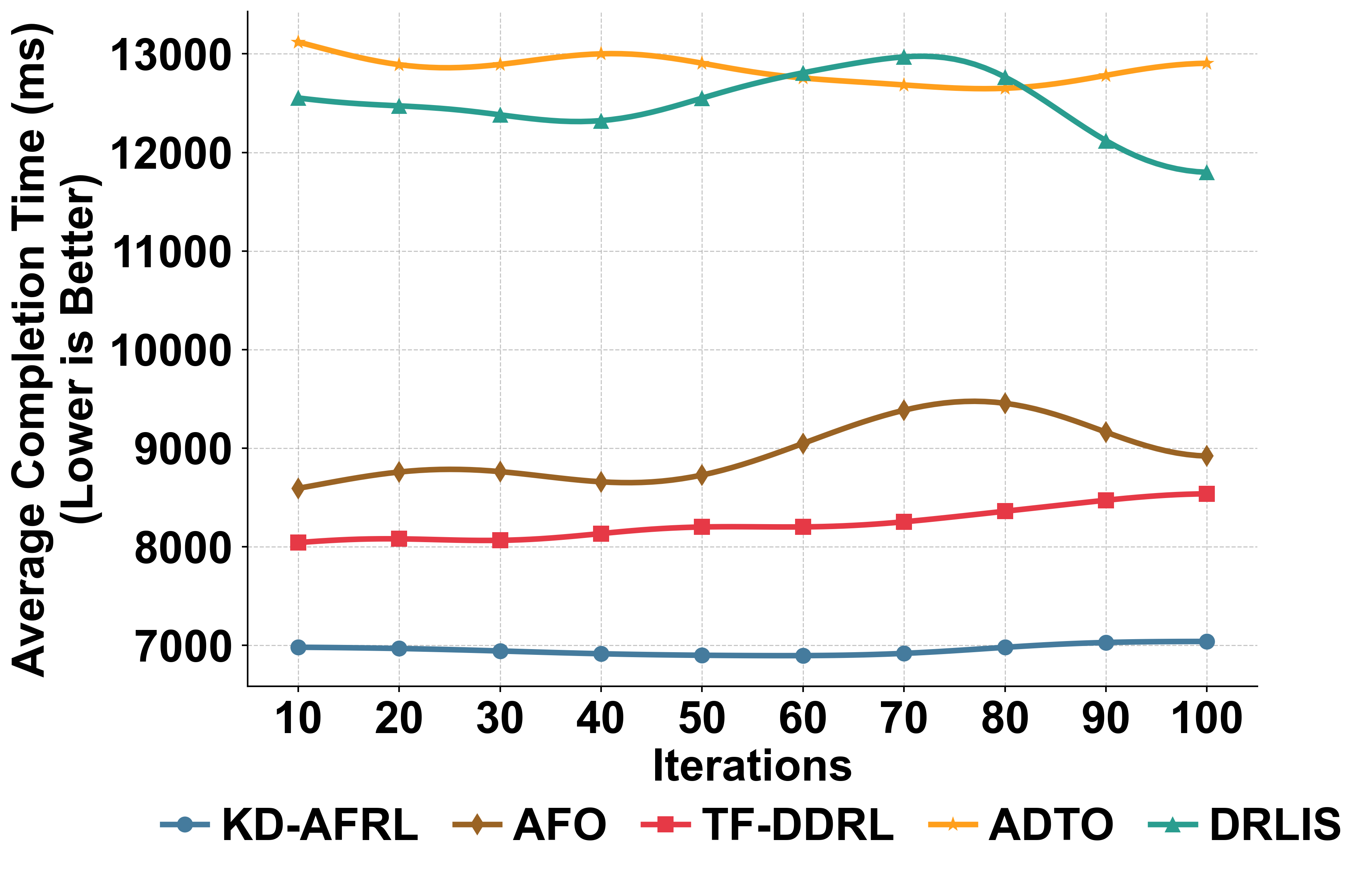}
  \caption{Average completion time}
  \label{fig:cte}
\end{subfigure}%
\begin{subfigure}{0.33\textwidth}
  \centering
  \includegraphics[width=\linewidth]{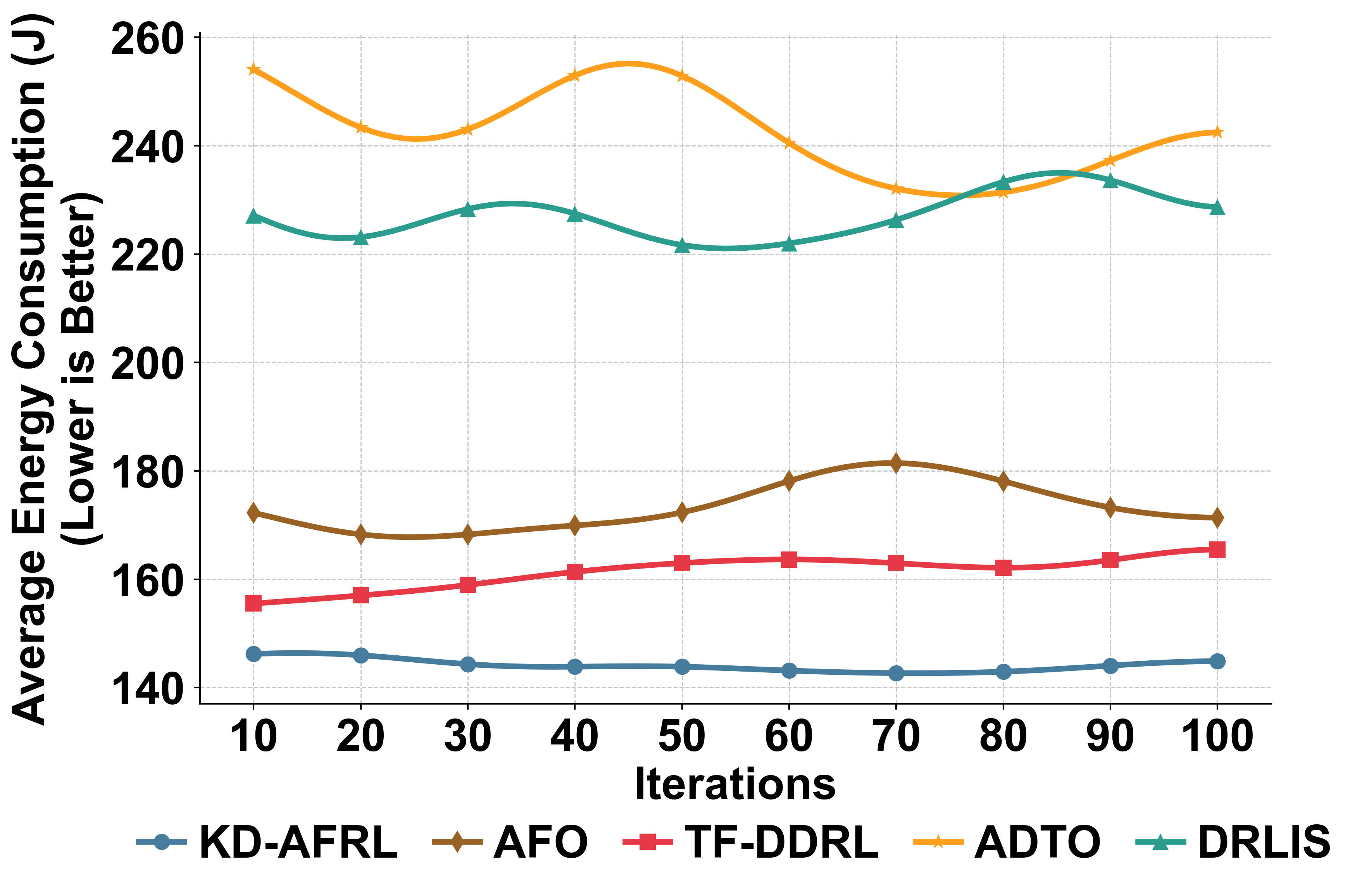}
  \caption{Average energy consumption}
  \label{fig:ece}
\end{subfigure}
\begin{subfigure}{0.33\textwidth}
  \centering
  \includegraphics[width=\linewidth]{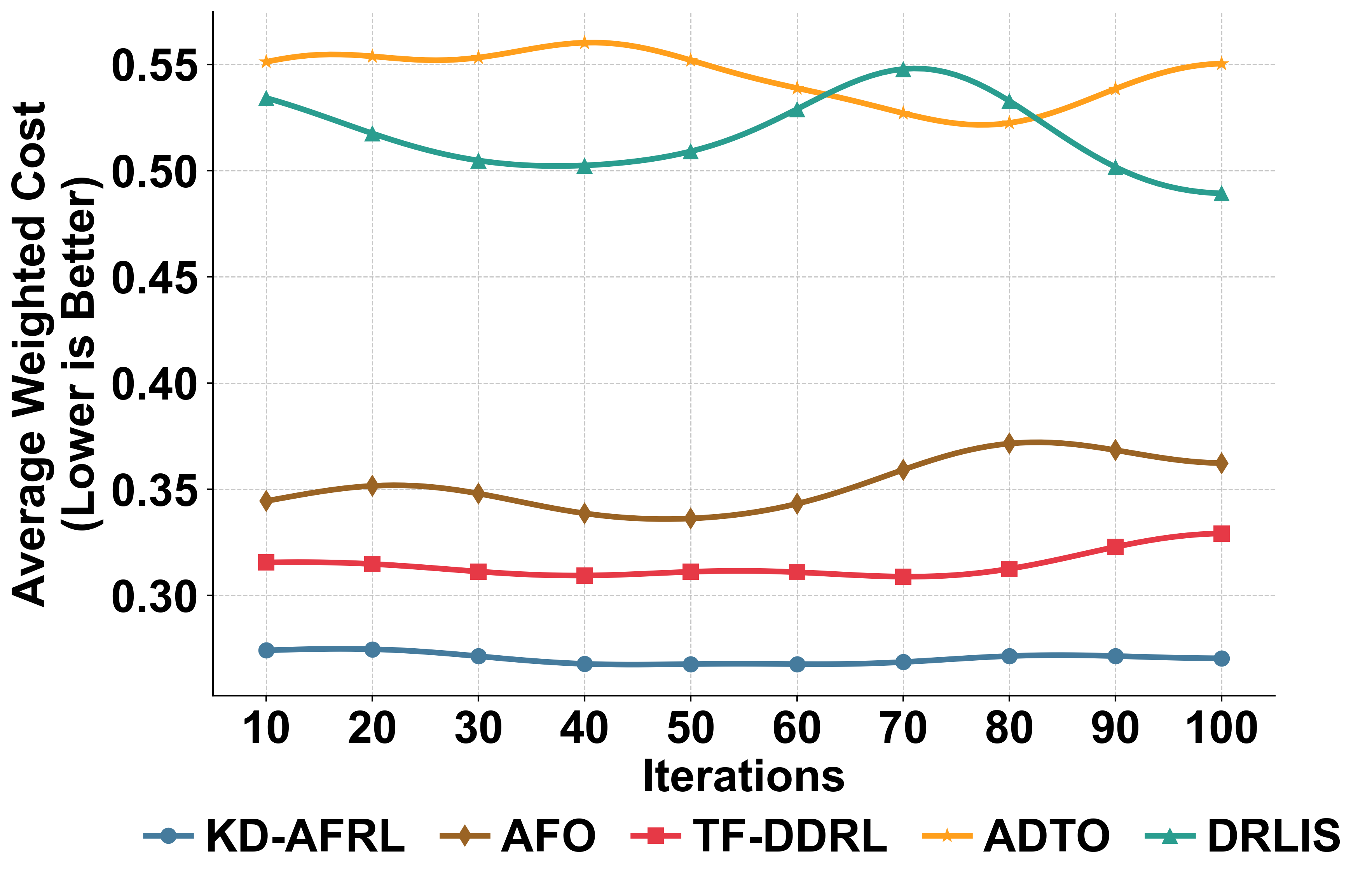}
  \caption{Average weighted cost}
  \label{fig:wce}
\end{subfigure}
\caption{Convergence performance analysis during evaluation phase}
\label{fig:cpa2}
\end{figure*}

\subsection{Hyperparameter Configuration}
We conducted systematic hyperparameter tuning using grid search with cross-validation for each domain. Learning rates and discount factors are optimized per scheduler. Table \ref{table:hyperparameters} summarizes the key hyperparameter settings. Baseline methods use identical tuning procedures.
\begin{table}[h]
\centering
\caption{The key hyperparameters setting for KD-AFRL}
\label{table:hyperparameters}
\footnotesize
\resizebox{\columnwidth}{!}{%
\begin{tabular}{ll|ll}
\hline
\textbf{Parameter} & \textbf{Value} & \textbf{Parameter} & \textbf{Value} \\
\hline
Learning Rate $\eta$ & [0.0001, 0.01] & Privacy Budget $\epsilon$ & 1.0 \\
Discount Factor $\gamma$ & [0.8, 0.99] & Drift Threshold $\tau_{drift}$ & 0.1 \\
Architecture Types $K_{arch}$ & 8 & Environmental Threshold $\tau_{env}$ & 0.6 \\
Network Layers & [2, 10] & Temperature $\tau_{temp}$ & 3.0 \\
Neurons per Layer & [16, 512] & Top-K Teachers $K$ & 3 \\
Cost Weight $\alpha_{cost}$ & 0.5 & Balancing Parameter $\alpha_{distill}$ & 0.7 \\
\hline
\end{tabular}%
}
\end{table}

\subsection{Experimental Results and Analysis}
The following subsections systematically analyze convergence performance, federated learning and knowledge distillation contributions, adaptive architecture effectiveness, and framework scalability.

\subsubsection{Convergence Performance Analysis}
To evaluate the learning efficiency and convergence characteristics of KD-AFRL, we design the experiment comprising training and evaluation phases. The training phase employs a dedicated set of training applications for policy learning and parameter updates, while the evaluation phase freezes the learning process and evaluates the generalization performance of trained policies using a completely different set of evaluation applications. This design ensures genuine assessment of generalization capabilities on unseen applications. To ensure fair comparison, all techniques are evaluated under identical conditions. The reported results represent the metrics averaged across all 10 scheduling domains.

Experimental results demonstrate KD-AFRL's superiority in both training efficiency and evaluation effectiveness. During the training phase (Fig. \ref{fig:cpa1}), KD-AFRL achieves optimal convergence within 70-80 iterations, approximately 21\% faster than TF-DDRL and AFO (90-100 iterations), while DRLIS and ADTO fail to converge within 100 iterations. In the evaluation phase (Fig. \ref{fig:cpa2}), KD-AFRL not only maintains stable performance on unseen applications but also outperforms the best baseline (TF-DDRL) by 15.7\%, 10.8\%, and 13.9\% in completion time, energy consumption, and weighted cost respectively, while all baseline techniques exhibit significant performance fluctuations and degradation. This superior performance stems from the synergistic effects of our three core mechanisms: adaptive architectures that prevent computational mismatches, environment clustering that ensures compatible domains share knowledge, and cross-architecture distillation that enables resource-constrained devices to learn from more capable ones. 

\subsubsection{Federated Learning and Knowledge Distillation Analysis}
\begin{figure}[h]
\begin{subfigure}{0.49\columnwidth}
  \centering
  \includegraphics[width=\linewidth]{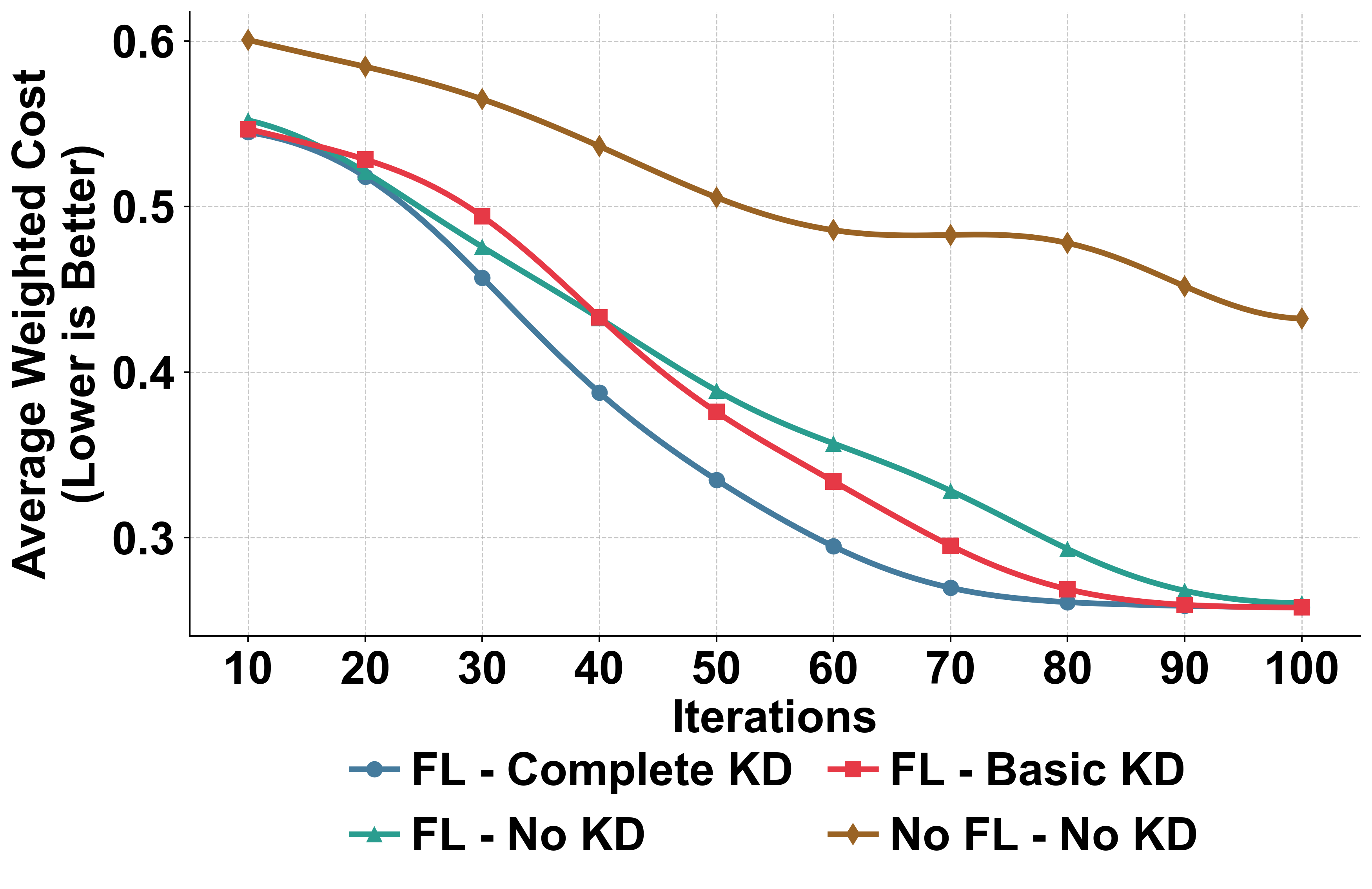}
  \caption{Training phase}
  \label{fig:kdt}
\end{subfigure}%
\begin{subfigure}{0.49\columnwidth}
  \centering
  \includegraphics[width=\linewidth]{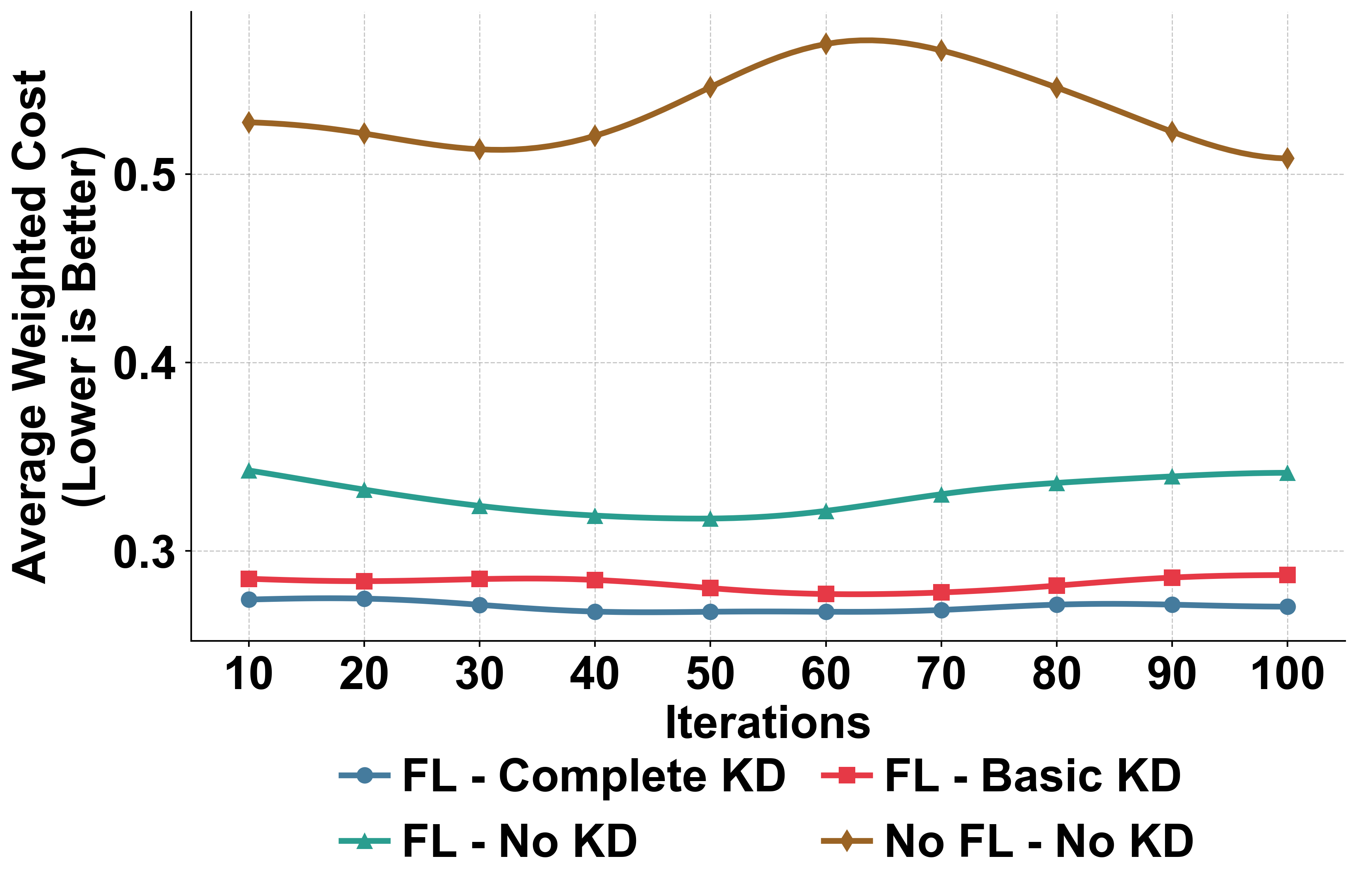}
  \caption{Evaluation phase}
  \label{fig:kde}
\end{subfigure}
\caption{Performance comparison of different learning strategies during training and evaluation phases}
\label{fig:kda1}
\end{figure}

To evaluate individual contributions of federated learning and knowledge distillation, we compare four learning strategies: No FL - No KD (local-training only), FL - No KD (federated only), FL - Basic KD (federated with basic distillation), and FL - Complete KD (federated with environment-oriented distillation). As shown in Fig.~\ref{fig:kda1}, FL - Complete KD achieves fastest convergence (80 iterations), outperforming FL - Basic KD (90 iterations), FL - No KD (100 iterations), and No FL - No KD (non-convergent). During evaluation, transitioning from local-only to federated learning reduces cost from 0.5 to 0.33 (34\% improvement), basic distillation further reduces it to 0.28, while complete distillation achieves optimal performance at 0.27. These results demonstrate that combining federated learning with environment-oriented knowledge distillation yields a 46\% overall improvement (from 0.5 to 0.27), where federated learning enables collaboration among compatible domains while knowledge distillation extends this benefit to heterogeneous models.

\begin{figure}[h]
\centering
\includegraphics[width=\linewidth]{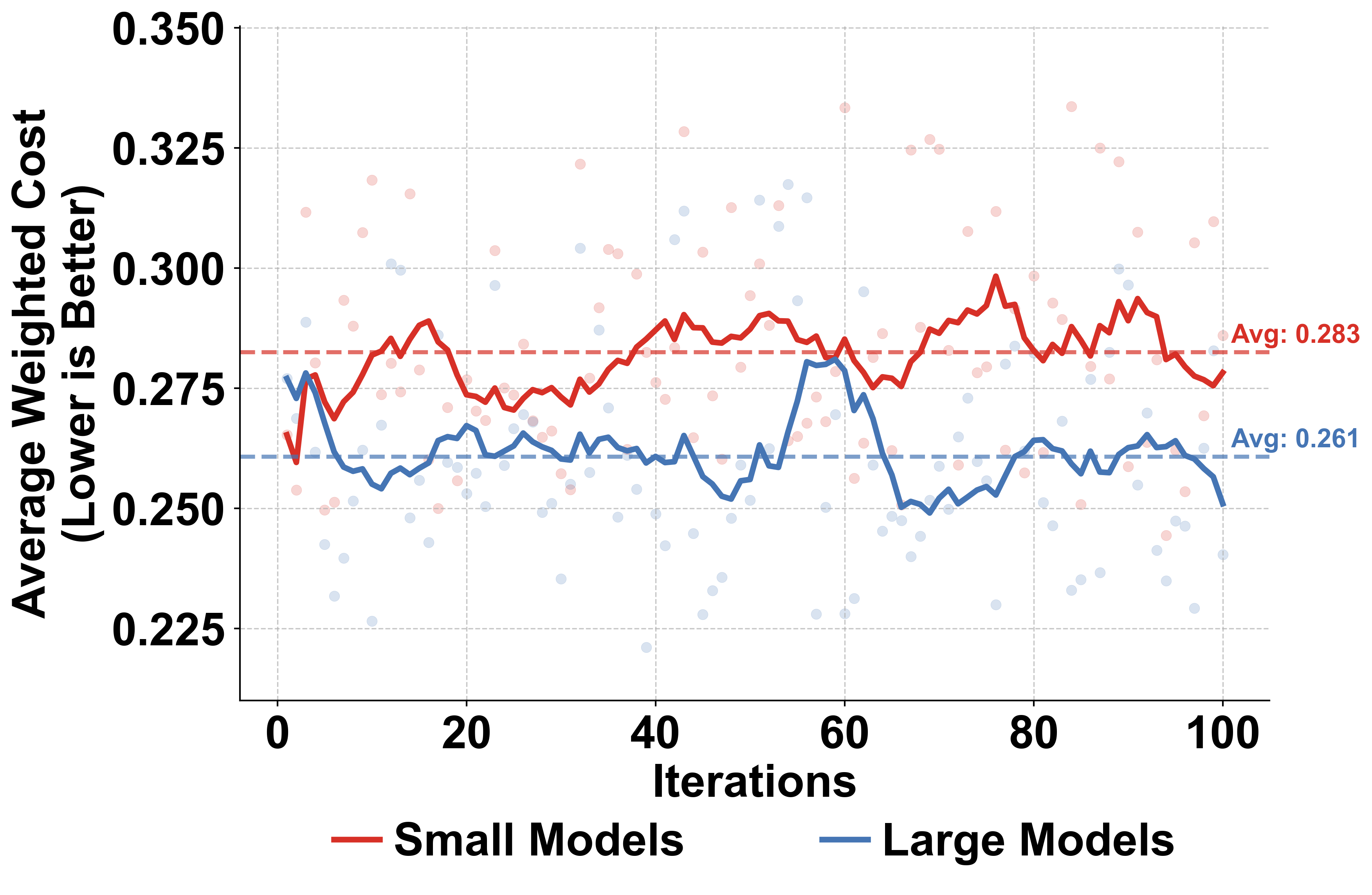}\vspace{-15pt}
\label{fig:kdc}
\caption{Cross-architecture knowledge distillation performance between small and large models}
\label{fig:kda2}
\end{figure}

To further demonstrate the effectiveness of cross-architecture knowledge distillation in bridging the performance gap between heterogeneous models, we analyze the performance differences between small models (2-4 layers with up to 32 neurons per layer) and large models (8-10 layers with up to 512 neurons per layer) under the FL - Complete KD strategy. As illustrated in Fig.~\ref{fig:kda2}, despite significant architectural disparities, the knowledge distillation mechanism successfully enables small models on resource-constrained devices to achieve an average weighted cost of 0.283, closely approaching the 0.261 achieved by large models on high-end devices. This represents 92.2\% relative efficiency, demonstrating that small models can effectively learn from and approximate the decision-making capabilities of their more complex counterparts.

\subsubsection{Adaptive Architecture Analysis}
To evaluate the effectiveness of the adaptive architecture generation mechanism, we conduct experiments in a heterogeneous distributed environment with multiple device types. Each scheduler is required to handle various types of applications with different computational demands. We deploy schedulers across three categories of heterogeneous devices, with multiple instances configured for each category: low-end devices (1-2 cores and 1-2 GB RAM), medium devices (4-8 cores and 8-16GB RAM), and high-end devices (8-32 cores and 32-128GB RAM). For comparison, we evaluate three approaches: (1) Fixed Small Model - a lightweight architecture with 2 hidden layers of 32 neurons deployed uniformly across all devices, (2) Fixed Large Model - a complex architecture with 8 hidden layers of 512 neurons deployed uniformly across all devices, and (3) Adaptive Model - our proposed approach that automatically adjusts architecture complexity based on device capabilities.

\begin{figure}[t]
\begin{subfigure}{0.49\columnwidth}
  \centering
  \includegraphics[width=\linewidth]{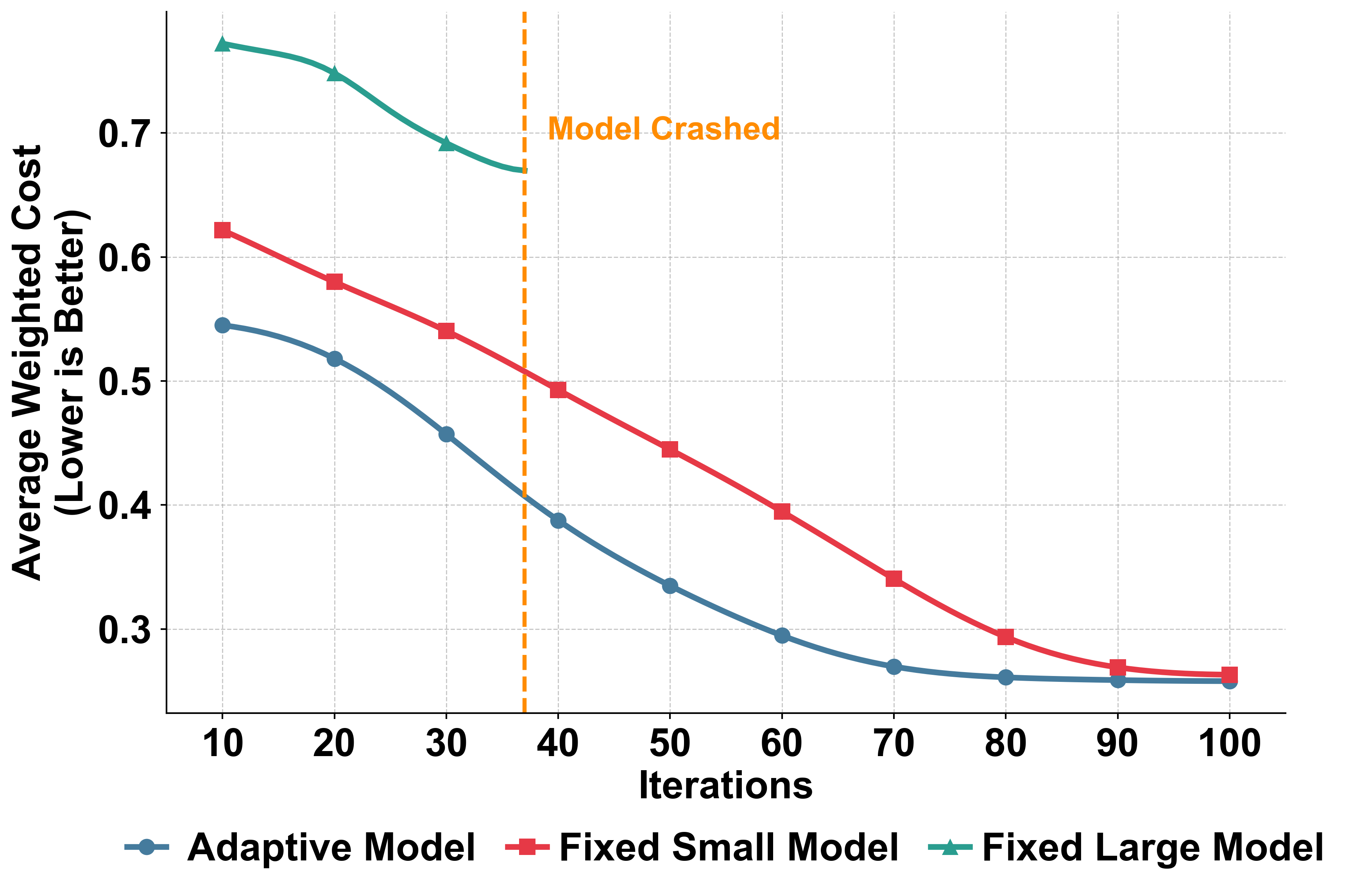}
  \caption{Training phase}
  \label{fig:aat}
\end{subfigure}%
\begin{subfigure}{0.49\columnwidth}
  \centering
  \includegraphics[width=\linewidth]{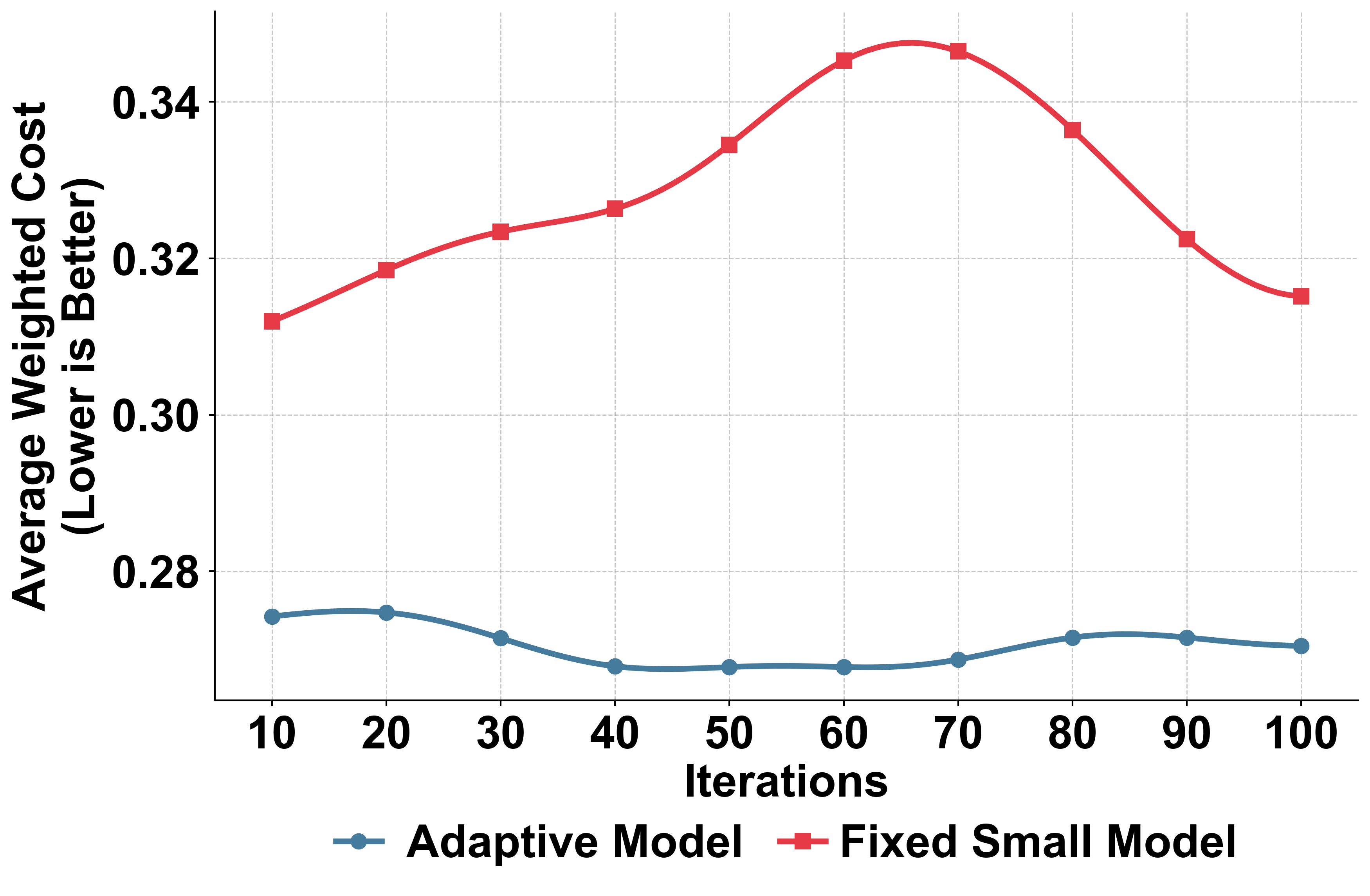}
  \caption{Evaluation phase}
  \label{fig:aae}
\end{subfigure}
\caption{Performance comparison of adaptive, fixed small, and fixed large models during training and evaluation phases}
\label{fig:aaa1}
\end{figure}

The training and evaluation phase results demonstrate the superior stability and convergence characteristics of the adaptive model. As shown in Fig.~\ref{fig:aaa1}, during training, the adaptive model achieves rapid convergence within approximately 80 iterations, while the fixed small model requires around 100 iterations, and the fixed large model crashes at iteration 37 due to resource constraints. During the evaluation phase, the adaptive model maintains stable optimal performance levels at around 0.27. In contrast, the fixed small model exhibits significant performance oscillations, with its weighted cost fluctuating between around 0.31 and 0.35.

\begin{figure}[t]
\begin{subfigure}{0.49\columnwidth}
  \centering
  \includegraphics[width=\linewidth]{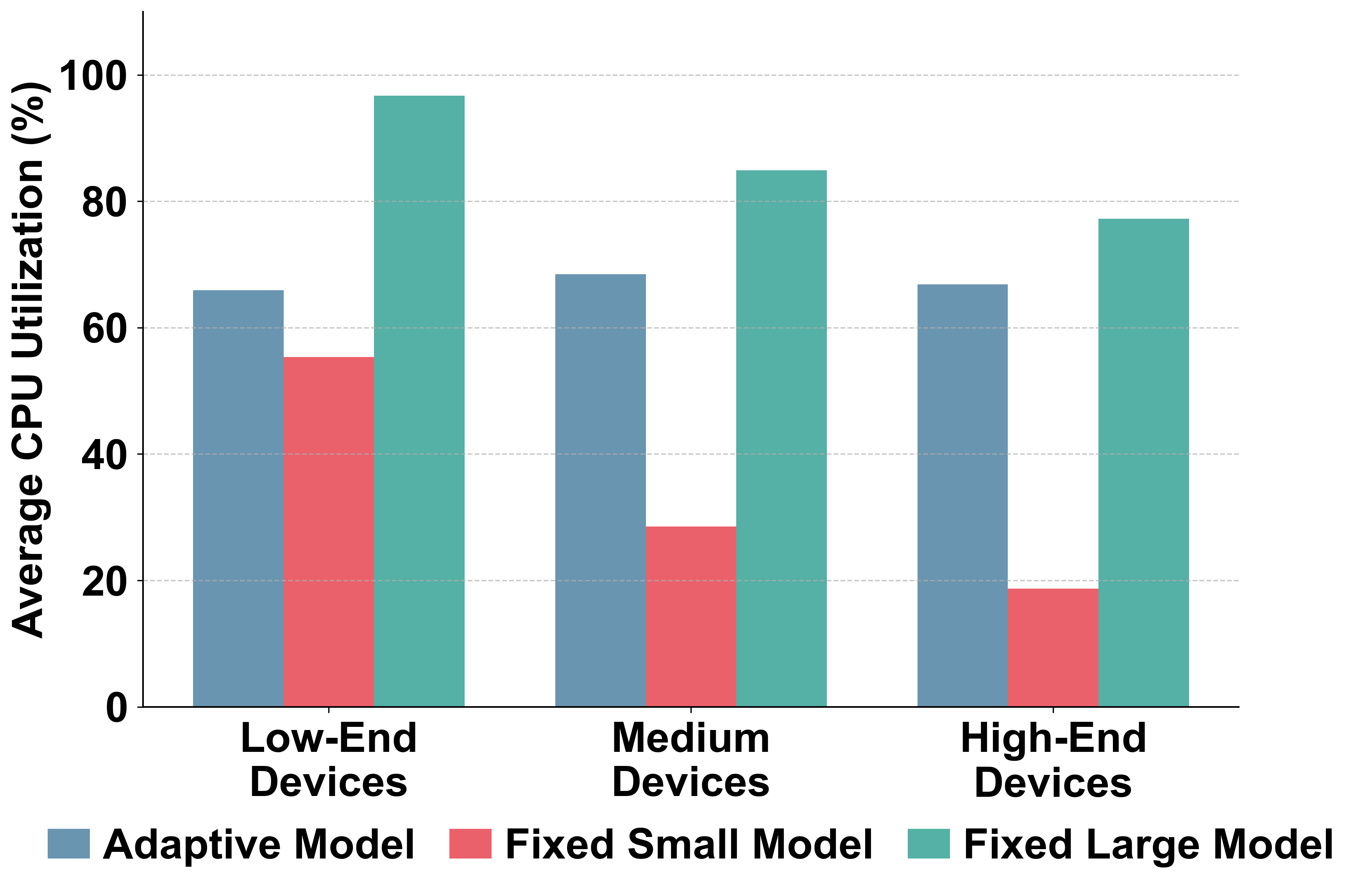}
  \caption{CPU resource utilization}
  \label{fig:cpu}
\end{subfigure}%
\begin{subfigure}{0.49\columnwidth}
  \centering
  \includegraphics[width=\linewidth]{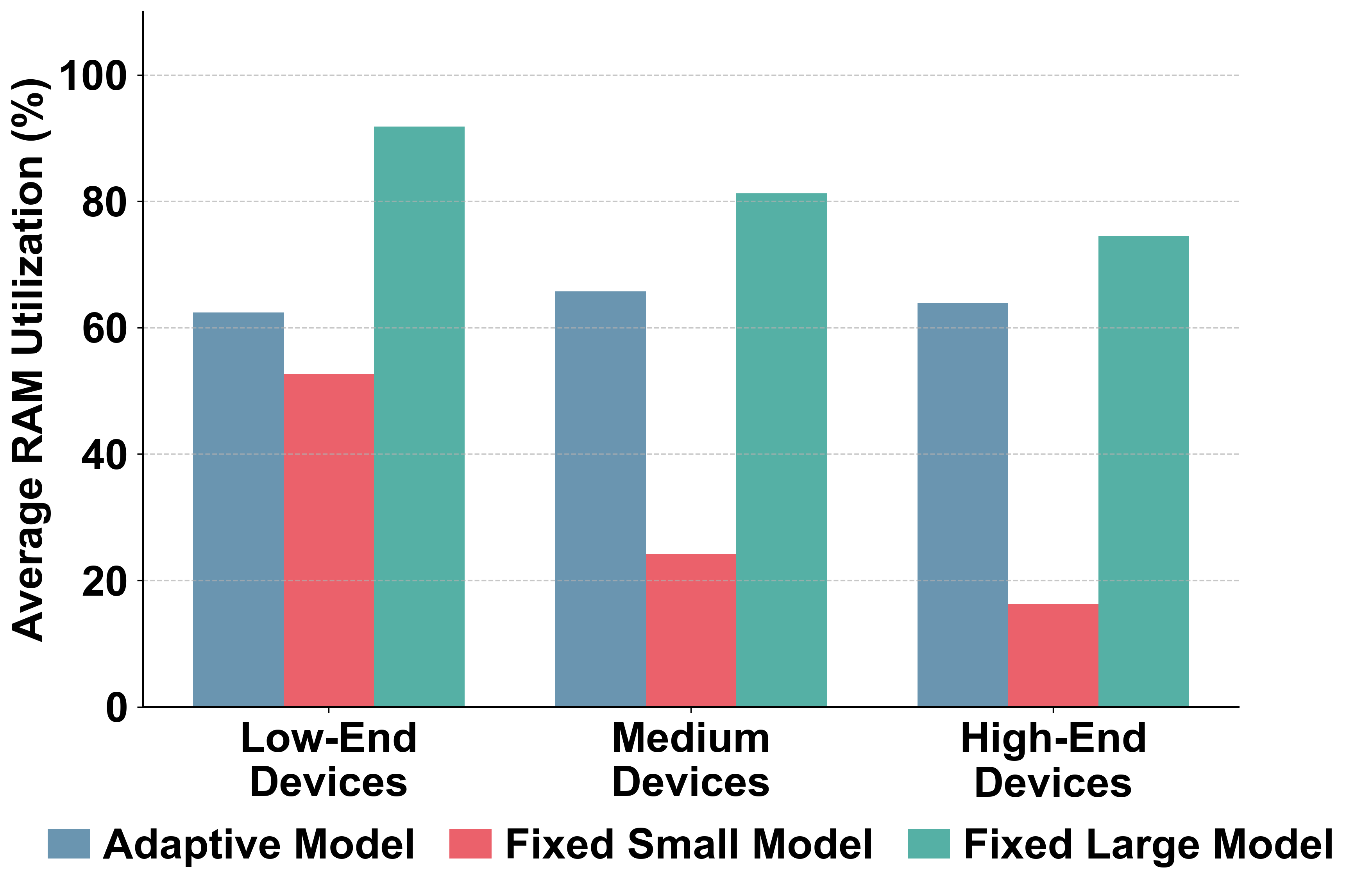}
  \caption{RAM resource utilization}
  \label{fig:ram}
\end{subfigure}
\caption{Resource utilization comparison across heterogeneous devices}
\label{fig:aaa2}
\end{figure}

We also examine the resource utilization patterns to further evaluate the effectiveness of the adaptive model. Fig.~\ref{fig:aaa2} shows the adaptive model achieves balanced resource utilization across all device types (CPU: 66-68\%, RAM: 62-66\%), while the fixed small model severely underutilizes resources on medium and high-end devices (CPU: 19-29\%, RAM: 16-24\%), and the fixed large model causes excessive consumption on low-end devices (CPU: 97\%, RAM: 92\%), leading to system failure. These results demonstrate that our adaptive architecture generation mechanism effectively balances computational demands and available resources in heterogeneous environments.

\begin{figure}[h]
\centering
\includegraphics[width=\linewidth]{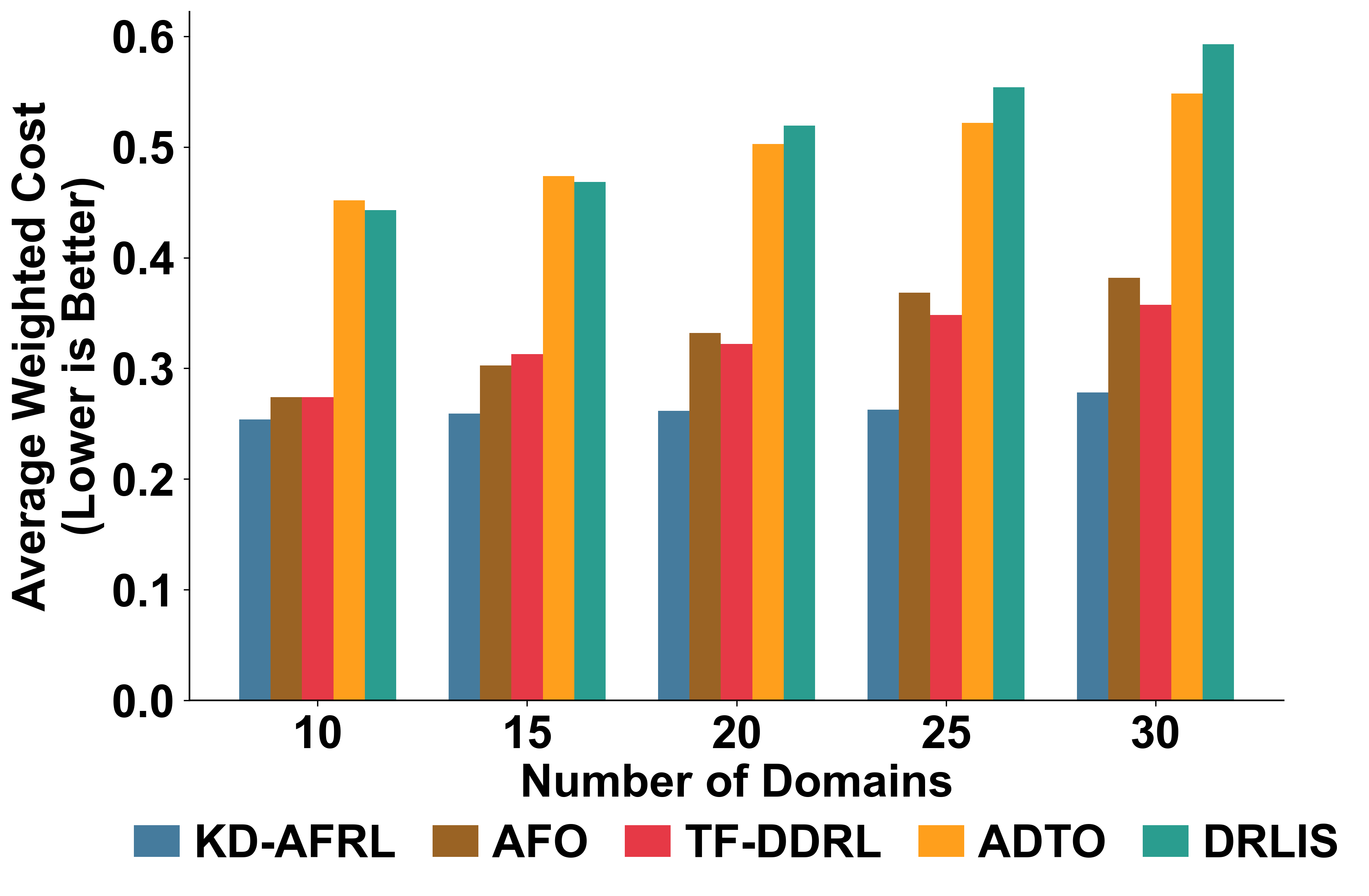}
\caption{Scalability performance comparison across varying number of domains}
\label{fig:sca}
\end{figure}

\subsubsection{Framework Scalability Analysis}
To evaluate the scalability of KD-AFRL, we design a dynamic expansion experiment that incrementally increases the number of domains from 10 to 30. As shown in Fig.~\ref{fig:sca}, KD-AFRL demonstrates superior scalability compared to all baseline methods across different domain sizes. When the number of domains increases from 10 to 30, KD-AFRL maintains the lowest average weighted cost, increasing from approximately 0.25 to 0.27, representing only an 8\% performance degradation. In contrast, the baseline techniques show significantly steeper performance decline: AFO increases from 0.27 to 0.38 (41\% degradation), TF-DDRL from 0.27 to 0.36 (33\% degradation), ADTO from 0.45 to 0.55 (22\% degradation), and DRLIS from 0.44 to 0.59 (34\% degradation). This demonstrates that KD-AFRL exhibits 3-5 times better performance retention compared to existing methods as the system scales. 

The superior scalability of KD-AFRL stems from the complementary effects of federated learning and environment-oriented knowledge distillation, enabling new domains to quickly learn from existing experienced domains without starting from scratch. This scalability advantage is crucial for practical deployments where IoT systems continuously expand with new domains joining.

\section{Conclusions and Future Work}
\label{conclusions}
In this paper, we propose KD-AFRL, a Knowledge Distillation-empowered Adaptive Federated Reinforcement Learning framework for multi-domain IoT application scheduling. The framework addresses existing limitations through three core innovations: resource-aware hybrid architecture generation, privacy-preserving environment-clustered federated learning, and environment-oriented cross-architecture knowledge distillation. Experimental evaluation demonstrates KD-AFRL achieves 21\% faster convergence and performance improvements of 15.7\%, 10.8\%, and 13.9\% in completion time, energy consumption, and weighted cost respectively, compared to the best baseline. Scalability experiments demonstrate that KD-AFRL achieves 3-5 times better performance retention compared to existing solutions as domains scale.

Future research directions include developing intelligent temperature regulation mechanisms for dynamic knowledge distillation optimization, exploring advanced privacy protection mechanisms such as homomorphic encryption and blockchain-based trust systems, and investigating automated neural architecture search techniques for dual-zone design optimization.

\ifCLASSOPTIONcaptionsoff
  \newpage
\fi

\bibliographystyle{IEEEtran}

\vfill

\end{document}